\documentclass[letterpaper,11pt]{article}
\pdfoutput=1
\usepackage{jheppub}
\usepackage{graphicx}
\usepackage{amsmath}
\usepackage{amssymb}
\usepackage{bm}
\usepackage{xspace}
\usepackage{snapshot}

\newcommand{\eq}[1]{eq.~\eqref{eq:#1}}

\newcommand{\eqs}[2]{eqs.~\eqref{eq:#1} and \eqref{eq:#2}}
\renewcommand{\sec}[1]{sec.~\ref{sec:#1}}

\newcommand{\fig}[1]{fig.~\ref{fig:#1}}

\newcommand{\ord}[1]{{\mathcal O}(#1)}
\newcommand{\ORd}[1]{{\mathcal O}\Bigl(#1\Bigr)}

\newcommand{\nn}{\nonumber}

\newcommand{\df}{\mathrm{d}}

\newcommand{\al}{\alpha}

\newcommand{\ga}{\gamma}

\newcommand{\de}{\delta}

\newcommand{\eps}{\epsilon}

\newcommand{\la}{\lambda}
\newcommand{\si}{\sigma}

\newcommand{\cG}{{\mathcal G}}

\newcommand{\lqcd}{\Lambda_\mathrm{QCD}}

\newcommand{\thr}{\mathrm{thr}}

\newcommand{\one}{{(1)}}

\newcommand{\Pythia}{\textsc{Pythia}\xspace}
\newcommand{\Herwig}{\textsc{Herwig}\xspace}
\newcommand{\Rivet}{\textsc{Rivet}\xspace}

\allowdisplaybreaks[2]

\preprint{\vbox{\hbox{Nikhef 2017-066}}}

\title{Phenomenology with a recoil-free jet axis: \\ TMD fragmentation and the jet shape}

\author[a]{Duff Neill,}
\author[b,c]{Andreas Papaefstathiou,}
\author[b,c]{Wouter J.~Waalewijn,}
\author[b,c]{Lorenzo Zoppi}

\affiliation[a]{Theoretical Division, MS B283, Los Alamos National Laboratory, Los Alamos, NM 87545, USA}
\affiliation[b]{Institute for Theoretical Physics Amsterdam and Delta Institute for Theoretical Physics, University of Amsterdam, Science Park 904, 1098 XH Amsterdam, The Netherlands}
\affiliation[c]{Nikhef, Theory Group, Science Park 105, 1098 XG, Amsterdam, The Netherlands}

\abstract{
We study the phenomenology of recoil-free jet axes using analytic calculations and Monte Carlo simulations. Our focus is on the average energy as function of the angle with the jet axis (the jet shape), and the energy and transverse momenta of hadrons in a jet (TMD fragmentation). We find that the dependence on the angle (or transverse momentum) is governed by a power law, in contrast to the double-logarithmic dependence for the standard jet axis. The effects of the jet radius, jet algorithm, angular resolution and grooming are investigated. TMD fragmentation is important for constraining the structure of the proton through semi-inclusive deep-inelastic scattering. These observables are also of interest to the LHC, for example to constrain $\alpha_s$ from precision jet measurements, or probe the quark-gluon plasma in heavy-ion collisions. 
}

\begin{document}
\maketitle

\section{Introduction}
\label{sec:intro}

\begin{figure}[b]
\centering
\includegraphics[width=0.5\textwidth]{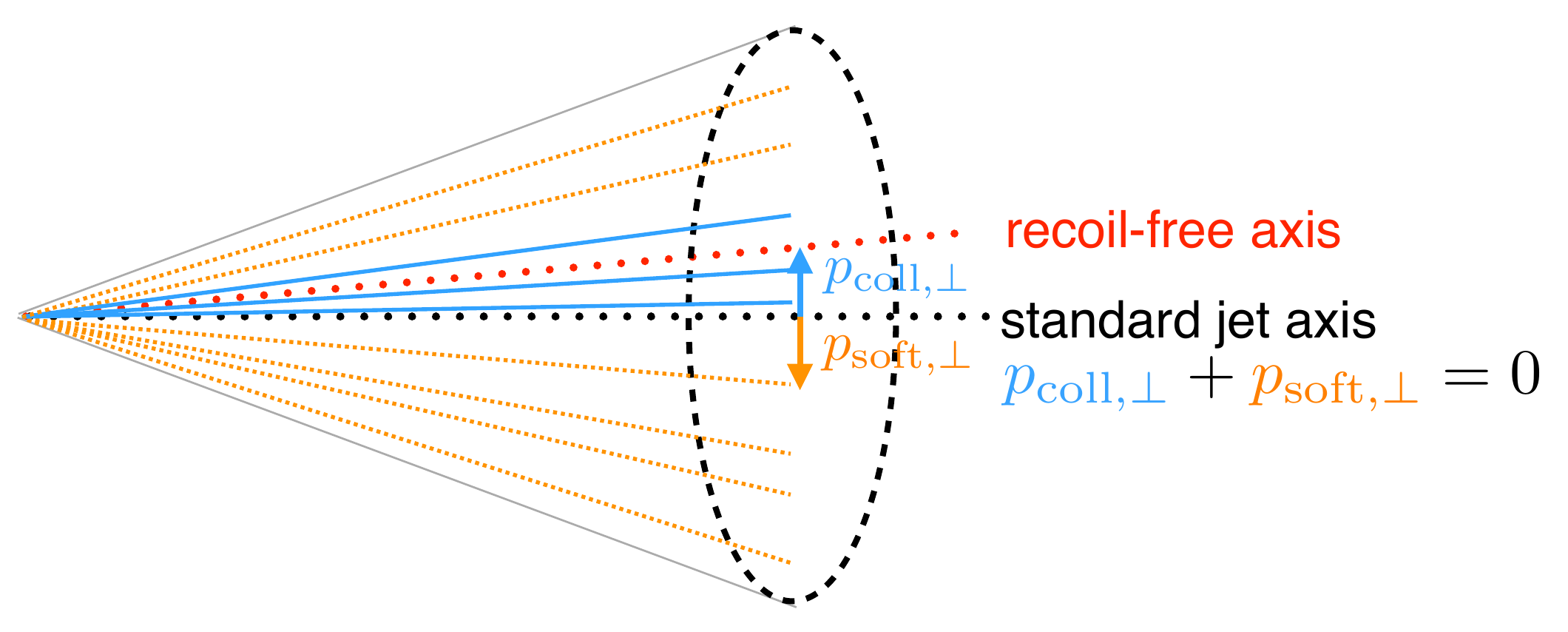}%
\caption{The standard jet axis (black) is along the total jet momentum, making it sensitive to the recoil of soft radiation (orange). By contrast, a recoil-free axis (red) tracks the energetic collinear radiation (blue).}
\label{fig:recoil}
\end{figure}

The jet axis is aligned with the total momentum of the particles inside the jet, for standard jet algorithms like Cambridge/Aachen \cite{Dokshitzer:1997in,Wobisch:1998wt,Wobisch:2000dk} and anti-$k_T$~\cite{Cacciari:2008gp}. This makes the direction of the axis sensitive to the recoil of soft radiation, see \fig{recoil}, which is undesirable for measuring the angular distribution of energetic radiation in a jet. Experimentally, it introduces an unnecessary sensitivity to soft radiation, causing the picture of the jet to be ``blurred" by all kinds of contamination. Theoretically, it complicates calculations due to e.g.~non-global logarithms~\cite{Dasgupta:2001sh}, that arise because the jet axis is sensitive to soft radiation inside but not outside the jet~\cite{Neill:2016vbi,Kang:2017mda}.

We avoid these complications by using a recoil-free axis, that follows the energetic radiation. Our main focus will be on the winner-take-all (WTA) axis~\cite{Salam:WTAUnpublished,Bertolini:2013iqa}, which modifies the recombination step in clustering algorithms. Specifically, the (massless) momenta $p_1^\mu = E_1(1, \hat n_1)$ and $p_2^\mu = E_2(1, \hat n_2)$ are recombined into the \emph{massless} momentum $p^\mu = (E_1+E_2) (1, \hat n)$, where $\hat n = \hat n_1$ if $E_1 > E_2$ and $\hat n = \hat n_2$ otherwise. From this definition it is clear that the effect of soft radiation is limited to its contribution to the energy, which is small. We will also discuss the broadening axis~\cite{Larkoski:2014uqa}, which is another recoil-free axis. However, we find that even at LL accuracy the calculation of its cross section is much more complicated than for the WTA axis. (Correspondingly, the broadening axis is also more difficult to implement in experimental studies.)

In this paper we perform a phenomenological study of the transverse momentum distribution (TMD) of energetic hadrons with respect to a recoil-free axis. The theory and phenomenology of fragmentation to hadrons in jets has been studied extensively~\cite{Procura:2009vm, Liu:2010ng, Jain:2011xz, Procura:2011aq,Jain:2012uq, Krohn:2012fg,Waalewijn:2012sv, Chang:2013rca, Chang:2013iba,Arleo:2013tya, Bauer:2013bza, Baumgart:2014upa, Ritzmann:2014mka,Kaufmann:2015hma,Chien:2015ctp,Bain:2016clc,Dai:2016hzf,Kang:2016ehg,Elder:2017bkd}, but the extension to TMDs was discussed only recently for the WTA axis~\cite{Neill:2016vbi} (and concurrently for the standard jet axis~\cite{Bain:2016rrv,Kang:2017glf,Kang:2017btw}).
The transverse momentum $\vec k_\perp$ is defined as
\begin{align} \label{eq:k_def}
  \vec k_\perp = \frac{\vec p_{h\perp}}{z_h} 
\,,\end{align}
where $z_h=E_h/E_J$ is the fraction of the jet energy carried by the hadron and $\vec p_{h\perp}$ its transverse momentum. This ensures that $\vec k_\perp$ is a partonic variable and thus calculable for $|\vec k_\perp|  \gg \lqcd$. We also show results using the angle $\theta$ between the hadron and the axis, which is related to $\vec k_\perp$ by
\begin{align} \label{eq:angle}
  k \equiv |\vec k_\perp| = E_J \sin \theta \approx E_J \theta 
\,.\end{align}
Much of the time we will consider the average energy due to all hadrons as function of $\theta$ or $k$, rather than considering the $z_h$ spectrum of an individual hadron species $h$. This corresponds to the jet shape~\cite{Ellis:1992qq,Seymour:1997kj,Li:2011hy,Chien:2014nsa,Kang:2017mda} but defined with respect to the WTA axis instead of the standard jet axis. 

To obtain analytic predictions, we use the formalism for TMD fragmentation with the WTA axis, that was recently developed by some of us in ref.~\cite{Neill:2016vbi}. We will compare our predictions to parton and hadron-level predictions obtained with \Pythia 8.2~\cite{Sjostrand:2014zea} and \Herwig 7.1~\cite{Bellm:2017bvx}, finding good agreement. We will also use Monte Carlo predictions to explore properties for which we did not (yet) perform an analytical calculation, such as grooming.

\begin{figure}[t]
\centering
\includegraphics[width=0.5\textwidth]{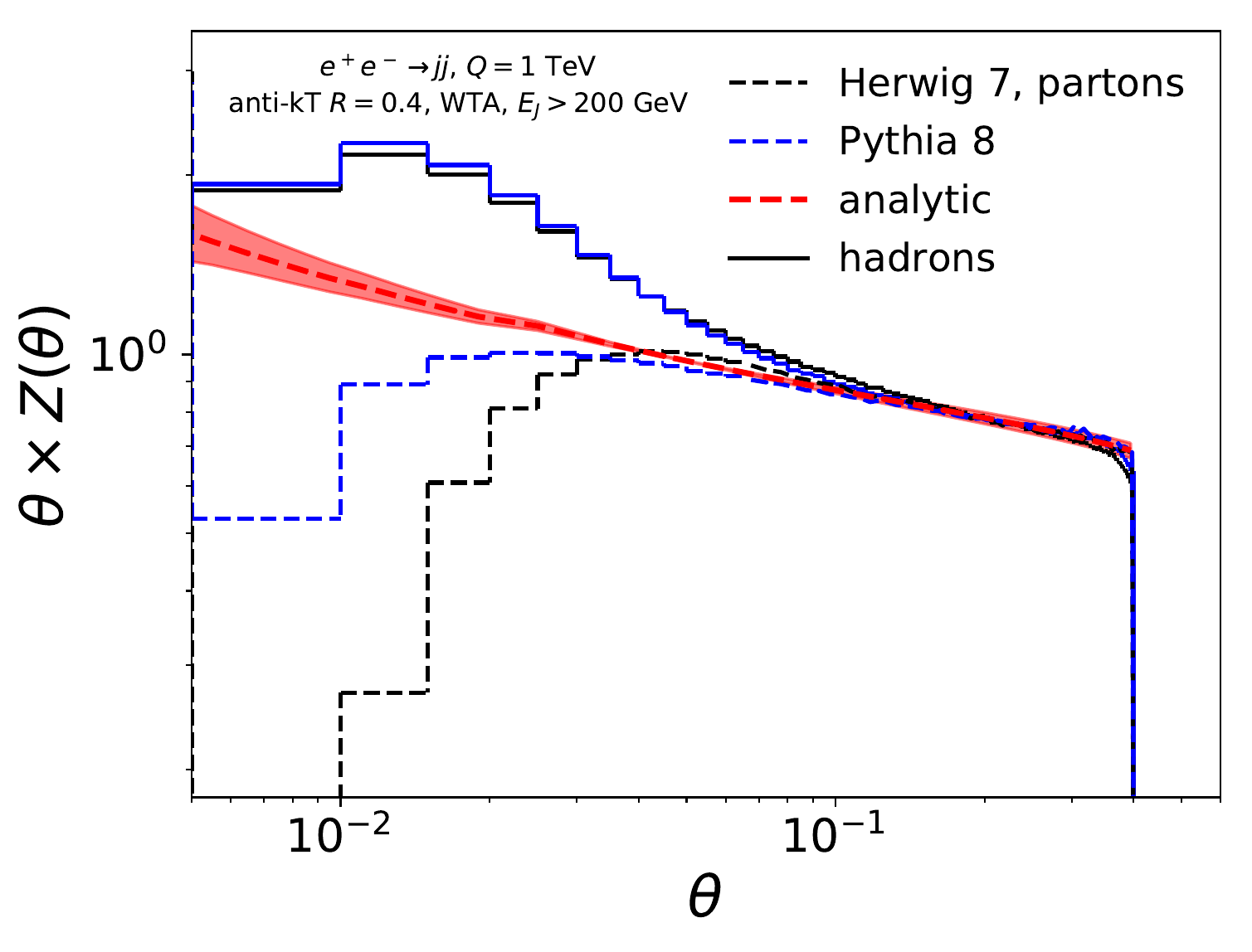}%
\caption{The energy distribution in a jet, see \eq{Z_def}, is a power law as function of the angle $\theta$ with the winner-take-all axis. Our analytic predictions agree with those obtained from \Pythia and \Herwig, except for very small angles where nonperturbative effects are important.}
\label{fig:moneyplot}
\end{figure}

Briefly highlighting some of our main findings: The dependence of the cross section on $k$ and $\theta$ is given by an (approximate) power law for the WTA axis, see \fig{moneyplot}, in contrast to the Sudakov double logarithms that appear for the standard jet axis. This remains true when taking into account the limited angular resolution, as long as $\theta$ is larger than the angular resolution scale. Furthermore, this also persists when restricting to charged particles, which allows one to exploit the finer angular resolution of the tracker. At small angles the distribution is rather sensitive to nonperturbative physics. For the standard jet axis these features are washed out due to smearing from soft radiation, so this provides a new opportunity to constrain nonperturbative collinear dynamics experimentally. The choice of jet algorithm is only visible in the region close to the jet boundary, $\theta = R$, as it affects which particles are inside or outside of the jet and not the axis. 

We envision several applications of our results: First of all, TMD fragmentation is of interest to the nuclear physics community in determining the structure of the proton, since the cross section in semi-inclusive deep-inelastic scattering involves a TMD parton distribution function and a TMD fragmentation function~\cite{Meng:1995yn}. Furthermore, these observables look promising for constraining $\alpha_s$, since they only involve collinear physics and can in principle be calculated to high orders in perturbation theory. As discussed above, they are also interesting for studying nonperturbative physics and could be used to improve hadronization models. Finally, the robustness of recoil-free axes makes them very interesting for probing the quark-gluon plasma through medium modifications.

The outline of this paper is as follows: In \sec{framework} we briefly review the framework for TMD fragmentation with the WTA axis in ref.~\cite{Neill:2016vbi}, rederiving the main results using a parton-shower picture. The broadening axis, threshold resummation, and details of our numerical implementation are also discussed. We present our results in \sec{results}, and conclude in \sec{conc}.

\section{Framework}
\label{sec:framework}

We start in \sec{TMDfrag} with reviewing the framework for TMD fragmentation with the winner-take-all axis of ref.~\cite{Neill:2016vbi}, addressing the factorization and resummation of logarithms arising from hierarchies between the center-of-mass energy $Q$, jet scale $E_J R$, transverse momentum $k=|\vec k_\perp|$ and the scale $\lqcd$ of nonperturbative physics. A short rederivation of these equations at leading logarithmic (LL) accuracy using a parton shower picture is presented in \sec{LL}.  In \sec{broadening} we investigate the broadening axis, finding that the corresponding cross section is much more complicated than for the WTA axis, even at leading logarithmic accuracy.  We then describe the connection between TMD fragmentation and the jet shape in \sec{jetshape}, discuss the resummation of threshold logarithms near the endpoint in \sec{threshold}, and provide details on our numerical implementation in \sec{numerical}.

\subsection{TMD fragmentation}
\label{sec:TMDfrag}

The cross section for producing a hadron $h$ with energy fraction $z_h$ and transverse momentum $\vec k_\perp = \vec p_{h\perp}/z_h$ inside a jet with energy $E_J$ and radius $R$ is given by 
\begin{align}\label{eq:match_si}
  \frac{\df \si_h}{\df E_J\, \df^2 \vec k_\perp\, \df z_h}
  = \sum_i \int\! \frac{\df x}{x}\, H_i\Big(\frac{E_J}{x}, \mu\Big)\, \cG_{i\to h}(x, E_J R, \vec k_\perp, z_h,\mu) \big[1 + \ord{R^2}\big]
\,.\end{align}
Here the collinear approximation $R \ll 1$ was exploited to factorize the cross section into a partonic cross section $H$, that describes the hard scattering, and the fragmenting jet function $\cG$, that captures the formation of the jet. Specifically, the parton $i$ with energy $E_J/x$ produced in the hard scattering emits radiation, resulting in a jet with energy $E_J$ that contains the hadron $h$. When the hadron is produced close to the center of the jet, i.e.~$k \equiv |\vec{k}_\perp| \ll E_J R$, we can furthermore separate the effects of the jet boundary $B$ from the fragmentation, 
\begin{align}\label{eq:match_G}
\cG_{i\to h}(x, E_J R, \vec k_\perp, z_h,\mu) = \sum_j \int\! \frac{\df y}{y}\, B_{ij}\Big(x, E_J R, \frac{z_h}{y},\mu\Big)\, D_{j \to h}(\vec k_\perp, y,\mu) \bigg[1 + \ORd{\frac{k^2}{E_J^2 R^2}} \bigg]
\,.\end{align}
This requires the effect of the measurement at angular scales $\theta \approx k/E_J$ and $R$ to factorize, which was argued to hold for the Cambridge/Aachen and anti-$k_T$ with the WTA axis in ref.~\cite{Neill:2016vbi}. Finally, for $k \gg \lqcd$, the transverse momentum dependence can be calculated,
\begin{align}\label{eq:match_D}
D_{j\to h}(\vec k_\perp, z_h,\mu) = \sum_k \int\! \frac{\df z}{z}\, C_{jk}\Big(\vec k_\perp, \frac{z_h}{z},\mu\Big)\, d_{k\to h}(z,\mu) \bigg[1 + \ORd{\frac{\lqcd^2}{k^2}} \bigg]
\,,\end{align}
where $d_{k \to h}(z,\mu)$ are the standard fragmentation functions~\cite{Georgi:1977mg,Ellis:1978ty,Collins:1981uw}.

We will also consider the case $E_J R \sim k \gg \lqcd$, where the hadron is produced fairly close to the jet boundary. The corresponding factorization theorem is given by
\begin{align}\label{eq:match_G_bis}
	\cG_{i\to h}(x, E_J R, \vec k_\perp, z_h,\mu) = \sum_k \int\! \frac{\df z}{z}\, J_{ik}\Big(x, E_J R, \vec k_\perp,\frac{z_h}{z},\mu\Big)\, d_{k \to h}(z,\mu) \bigg[1 + \ORd{\frac{\lqcd^2}{E_J^2R^2}} \bigg].
\end{align}
The coefficients $J_{ik}$ were also computed in ref.~\cite{Neill:2016vbi}. When $k \ll E_JR$, they factorize as
\begin{align}\label{eq:match_J}
  J_{ik}(x, E_J R, \vec k_\perp,z,\mu) = \sum_j \int\frac{\df z'}{z'} B_{ij}\Big(x, E_J R, \frac{z}{z'},\mu\Big) C_{jk}(\vec{k}_\perp,z',\mu)\bigg[1 + \ORd{\frac{k^2}{E_J^2 R^2}} \bigg],
\end{align}
as required by consistency with \eqs{match_G}{match_D}. By using $J_{ik}$ when $k \sim E_J R$, instead of the factorized form on the right-hand side, we capture the $k^2/(E_J^2 R^2)$ corrections that are crucial in this region.

The logarithms in the cross section in \eq{match_si} become large when there are hierarchies between the scales of the hard scattering $Q$, the jet energy $E_J R$ and the transverse momentum $\vec k_\perp$. They can be resummed by evaluating each ingredient at its natural scale
\begin{align} \label{eq:canonical}
  \mu_H \sim Q\,, \qquad
  \mu_B \sim E_J R\,, \qquad
  \mu_C \sim k\,,
\end{align}
and using the renormalization group evolution (RGE) to evolve them to a common scale. The RGEs are~\cite{Kang:2016mcy,Dai:2016hzf,Neill:2016vbi} 
\begin{align}  \label{eq:RGE}
  \frac{\df} {\df\ln \mu}\, \cG_{i\to h}(x, E_J R, \vec k_\perp, z_h,\mu) &= \sum_j \int \frac{\df x'}{x'}\, \ga_{ij}\Big(\frac{x}{x'},\mu\Big)\, \cG_{j \to h}(x', E_J R, \vec k_\perp, z_h,\mu) 
  \,, \nn \\
  \frac{\df} {\df\ln \mu}\, D_{i\to h}(\vec k_\perp, z_h,\mu) 
  & = \sum_j \int \frac{\df z}{z}\, \ga_{ij}'\Big(\frac{z_h}{z},\mu\Big)\, D_{j\to h}(\vec k_\perp, z,\mu)
  \,,  \nn \\  
 \frac{\df} {\df\ln \mu}\, d_{i\to h}(z_h,\mu) 
  &= \sum_j \int \frac{\df z}{z}\, \ga_{ij}\Big(\frac{z_h}{z}, \mu\Big) d_{j\to h}(z, \mu)
\,,\end{align}
with anomalous dimensions
\begin{align} \label{eq:gamma}
  \ga_{ij}(z,\mu) &= P_{ji}(z,\mu)
  \,,\nn \\ 
  \ga_{ij}'(z,\mu) &= \theta\Big(z \geq \frac12\Big)\, P_{ji}(z,\mu)
\,.\end{align}
Here $P$ denote the time-like DGLAP splitting functions~\cite{Gribov:1972ri,Altarelli:1977zs,Dokshitzer:1977sg}. The $\theta(z \geq \tfrac12)$ is due to the winner-take-all axis, and will be rederived at LL in the next section. Details on our numerical implementation of the above equations are given in \sec{numerical}.

\subsection{Leading-logarithmic derivation}
\label{sec:LL}

\begin{figure}[t]
\centering
\includegraphics[width=0.46\textwidth]{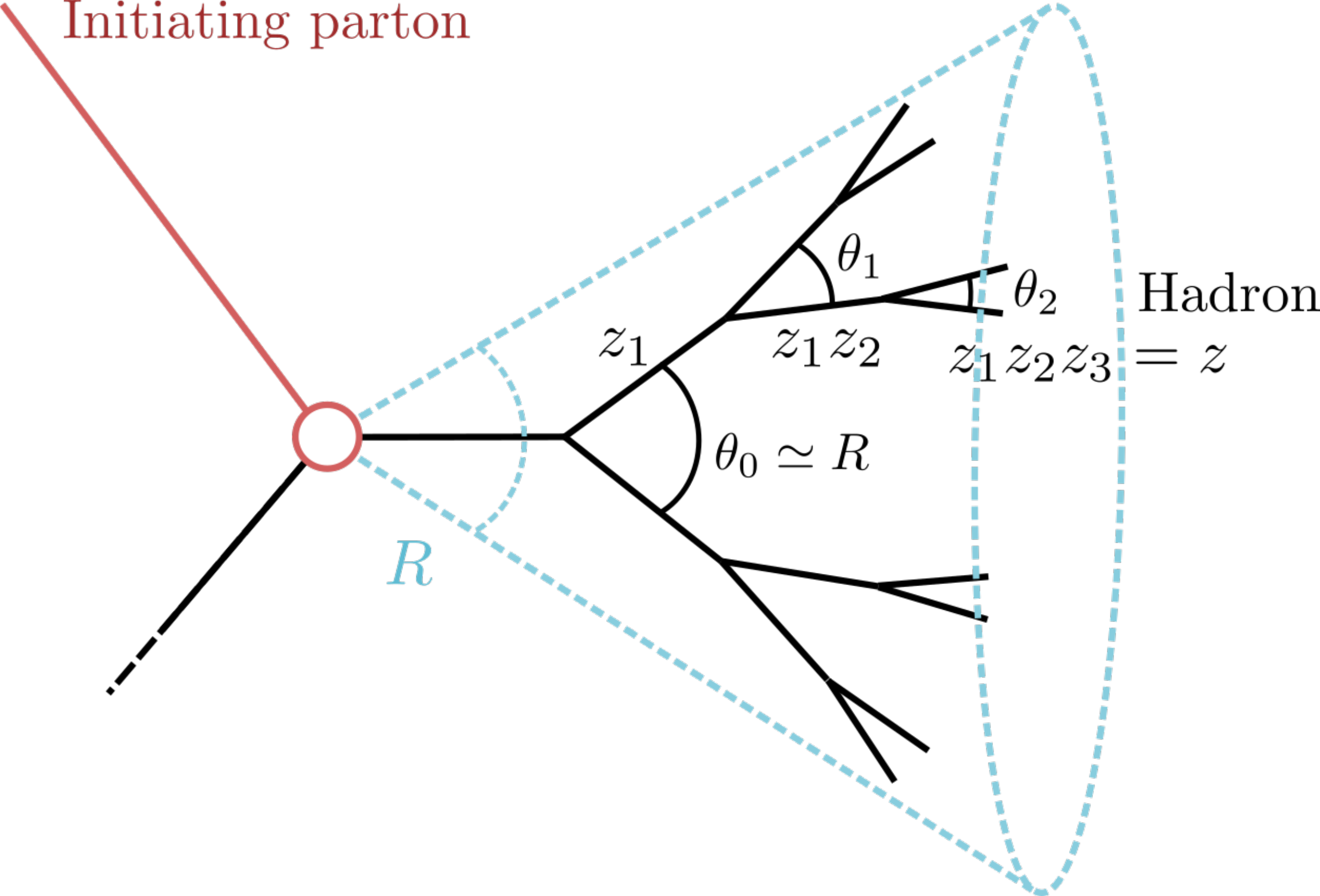}%
\caption{The parton shower picture for fragmentation in a jet. At LL accuracy the splittings are strongly ordered in angle, causing the shower tree and clustering tree of the jet algorithm to coincide.}
\label{fig:shower_tree}
\end{figure}

We now rederive the results of ref.~\cite{Neill:2016vbi}, which we summarized in \sec{TMDfrag}. We will use a parton shower picture which is valid to LL accuracy. Specifically, the radiation emitted by the parton produced in the hard interaction is described by a binary tree, see \fig{shower_tree}, where each mother splits into two daughters with (relative) momentum fractions $z_i$ and $1-z_i$ and angle $\theta_i$ between them. The tree is angular ordered, i.e.~angles of subsequent emissions are (parametrically) smaller. 

Let us first consider fragmentation in a jet before addressing TMD fragmentation with the WTA axis. At LL order, the corresponding fragmenting jet function (which differs from $\mathcal{G}$ in \eq{match_si}) reduces to the fragmentation function
\begin{align} \label{eq:ex}
  \cG_{i\to h}(x, E_J R,  z_h,\mu) &= \de(1-x)\, d_{i \to h}(z_h, \mu) 
\,,\end{align}
evaluated at $\mu \sim E_J R$ to minimize higher-order corrections. In terms of a parton shower this would be described by
\begin{align} \label{eq:LL_ex}
  A &= 
  d_h(z_h,R_0) \!+\!  \sum_{n=1}^{\infty} \bigg( \prod_{i=1}^n \int_0^1\! \df z_i\, P(z_i) \! \int_{R_0}^{\theta_{i-1}}\! \frac{\df \theta_i}{\theta_i} \bigg) 
  \! \int_0^1\! \df z'\, d_h(z',R_0)\, \de\Big(z_h \!-\! z' \prod_{j=1}^n\, z_j\Big)  
\,.\end{align}
To keep the discussion simple, we have ignored parton flavors. The first term corresponds to the case where the initial parton does not undergo any perturbative splitting. We then sum over $n$ emissions, and for each emission integrate over its splitting fraction $z_i$ and angle $\theta_i$, with a probability described by the splitting function $P$. This  follows from a repeated application of the collinear approximation, with splitting probability
\begin{align} \label{eq:coll_approx}
   \int_0^1\! \df z_i\, P(z_i) \! \int \frac{\df \theta_i}{\theta_i} 
\,.\end{align}
The upper bound on the $\theta_i$ integration follows from the angular ordering, which for the angle $\theta_1$ of the first splitting is the jet radius $R$. We regulate the collinear singularity at small angles with a cutoff $R_0$ and describe subsequent (nonperturbative) splittings by the fragmentation function $d_h$. 
Note that instead of traversing all branches of the binary tree, we follow the branch with splitting fraction $z_i$ (rather than $1-z_i$), since the other branches are effectively included because we integrate over $z_i$. The observed momentum fraction is the product of all $z_i$ and $z'$, as described by the measurement delta function. 

We now show that \eq{LL_ex} reproduces \eq{ex}. The angular integrals simply yield
\begin{align}
   \prod_{i=1}^n \int_{R_0}^{\theta_{i-1}}\! \frac{\df \theta_i}{\theta_i} = \frac{1}{n!} \ln^n \frac{R}{R_0}
\,.\end{align}
Since $\mu = E_J R$ in \eq{ex}, we can use $\df \ln \mu = \df \ln R$ to obtain,
\begin{align} \label{eq:DGLAP_ex}
 \frac{\df A}{\df \ln \mu} 
  &= 
  \int_0^1\! \df z\,
  \sum_{n=1}^{\infty} 
  \bigg( \prod_{i=1}^{n-1} \int_0^1\! \df z_i\, P(z_i)\bigg)\, \frac{1}{(n-1)!} \ln^{n-1} \frac{R}{R_0}\,  \int_0^1\! \df z'\, d_h(z',R_0)\, \de\Big(z - z' \prod_{j=1}^{n-1}\, z_j\Big)
  \nn \\ & \quad \times
  \int_0^1\! \df z_n\, P(z_n)\, \de(z_h- z z_n)
  \nn \\
  &= \int_{z_h}^1 \frac{\df z}{z} P\Big(\frac{z_h}{z}\Big) A(z)
\,,\end{align}
where on the first line we separated out the integral over $z_n$ to recognize the remainder as $A$.
This shows that $A$ satisfies the DGLAP evolution. Furthermore for $R=R_0$, $A$ is simply equal to the fragmentation function, so the boundary condition is also correct.

\begin{figure}[t]
\centering
\includegraphics[width=0.5\textwidth]{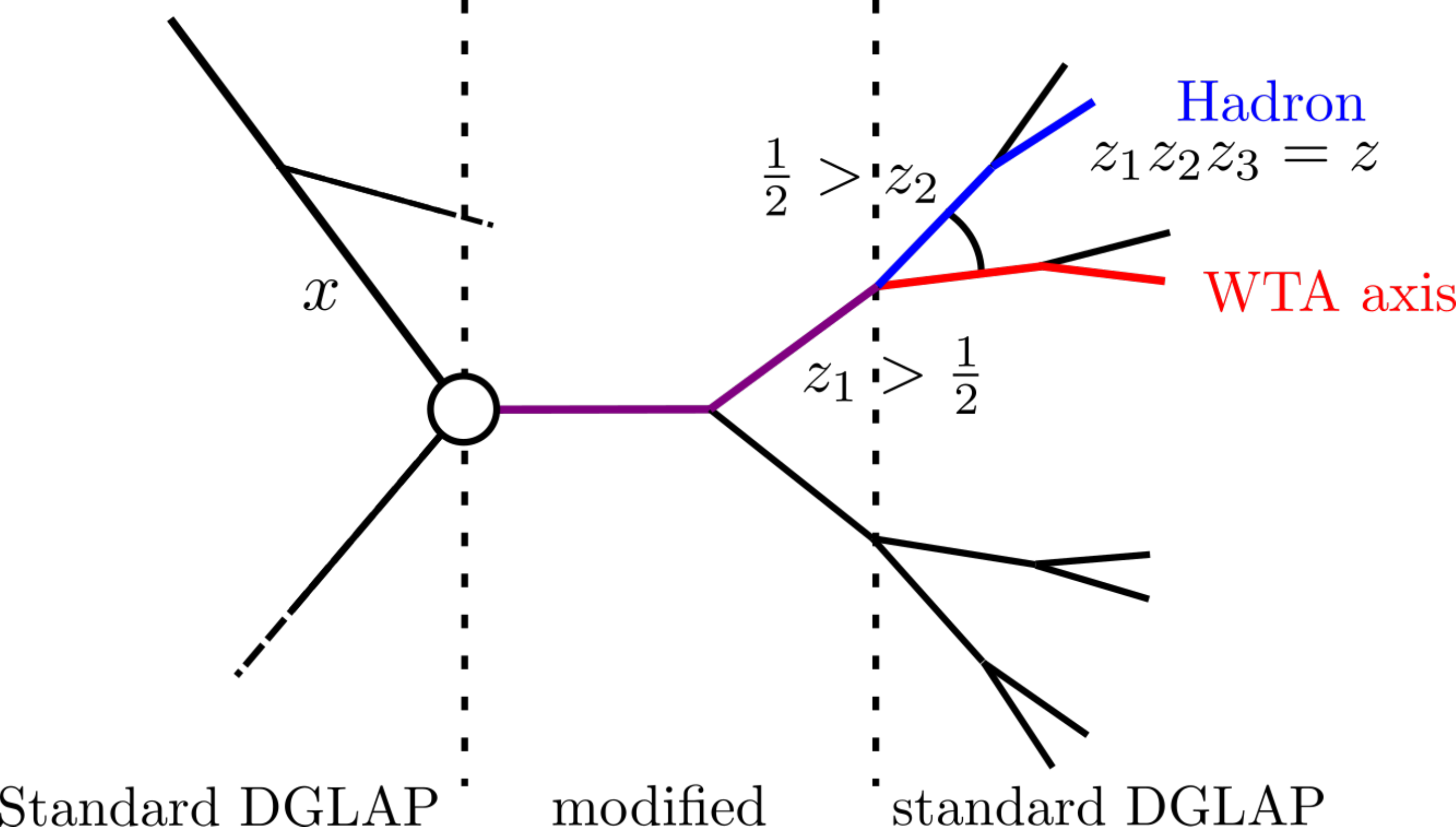}%
\caption{The parton shower picture for TMD fragmentation with the WTA axis. At each splitting the axis (purple/red) follows the largest momentum fraction. The angular ordering implies that the hadron must follow the same branch (purple) until the splitting angle drops below $k/E_J$, after which it can take a different branch (blue).}
\label{fig:WTA_TMDtree}
\end{figure}

We now extend this to TMD fragmentation with the winner-take-all axis. We can split the parton shower into three segments, see \fig{WTA_TMDtree}:
\begin{enumerate}
\item[(a)] $\theta_i>R$: All branches yield separate jets and are summed over, since we consider an inclusive jet sample. Splittings modify $x = 2E_J/Q$ but don't affect $z_h$ because it is defined relative to the jet energy.

\item[(b)] $R \geq \theta_i > k/{\rm E_J}$: Splittings take place inside the jet and thus do not modify $x$ but will affect $z_h$. Because $\theta_i > k/{\rm E_J}$ and the strong ordering in angles, the WTA axis and fragmenting hadron must at this point in the shower still follow the same branch.

\item[(c)] $k/{\rm E_J} \geq \theta_i$: The first splitting sets the angle (or equivalently $\vec k_\perp$) between the WTA axis and the hadron. Subsequent splittings modify $z_h$ but cannot change this angle due to the strong angular ordering. All of these emissions are summed over.
\end{enumerate}
We can directly repeat the above parton shower analysis in \eqs{LL_ex}{DGLAP_ex}, from which it follows that that in (a) there is a DGLAP evolution in $x$ from $\mu = Q$ down to $\mu = E_J R$, and in (c) there is a DGLAP evolution in $z_h$ from $\mu = k$  to the initial scale $\mu \sim \lqcd$ of the fragmentation functions. For (b) we note that strong ordering in angles implies that the clustering tree of any jet algorithm of the $k_T$ family (not just Cambridge/Aachen and anti-$k_T$) coincides with the parton shower tree. For $R \geq \theta_i > k/{\rm E_J}$  we only follow the branch that will yield the winner-take-all axis and produce the hadron, which corresponds to imposing $z_i>\tfrac12$, leading to the modified anomalous dimension $\ga'$ in \eq{gamma}. Thus we have justified the evolution equations in \eqs{RGE}{gamma} and the scales in \eq{canonical}.

\subsection{Broadening axis}
\label{sec:broadening}

We now investigate the broadening axis, which is the other well-known recoil-free jet axis. It is defined by minimizing the broadening, i.e.~it is the unit-vector $\hat n$ that minimizes the scalar sum of transverse momenta in the jet~\cite{Larkoski:2014uqa}.
\begin{align}
 b = \underset{\hat{n}}{\mbox{min}} \,(b_{\hat{n}}) 
 \,, \qquad 
 b_{\hat{n}} = \frac{1}{E_J} \sum_{i \in {\rm jet}} E_i \Big|
 2 \sin \frac{\vartheta_{i,\hat{n}}}{2} \Big| \,\approx\, \sum_{i\in {\rm jet}} z_i\, |\vartheta_{i,\hat{n}} |
\,.\end{align}
When the jet consists of two particles, the broadening axis is along the most energetic one, just as the WTA axis. However, we will show that the resummation of logarithms of $k/E_J R$ takes on a much more complicated form than for the WTA axis, because the axis finding does not have a simple recursive picture as for the WTA in \sec{LL}, even at LL accuracy. 

\begin{figure}[t]
\centering
\includegraphics[width=0.3\textwidth]{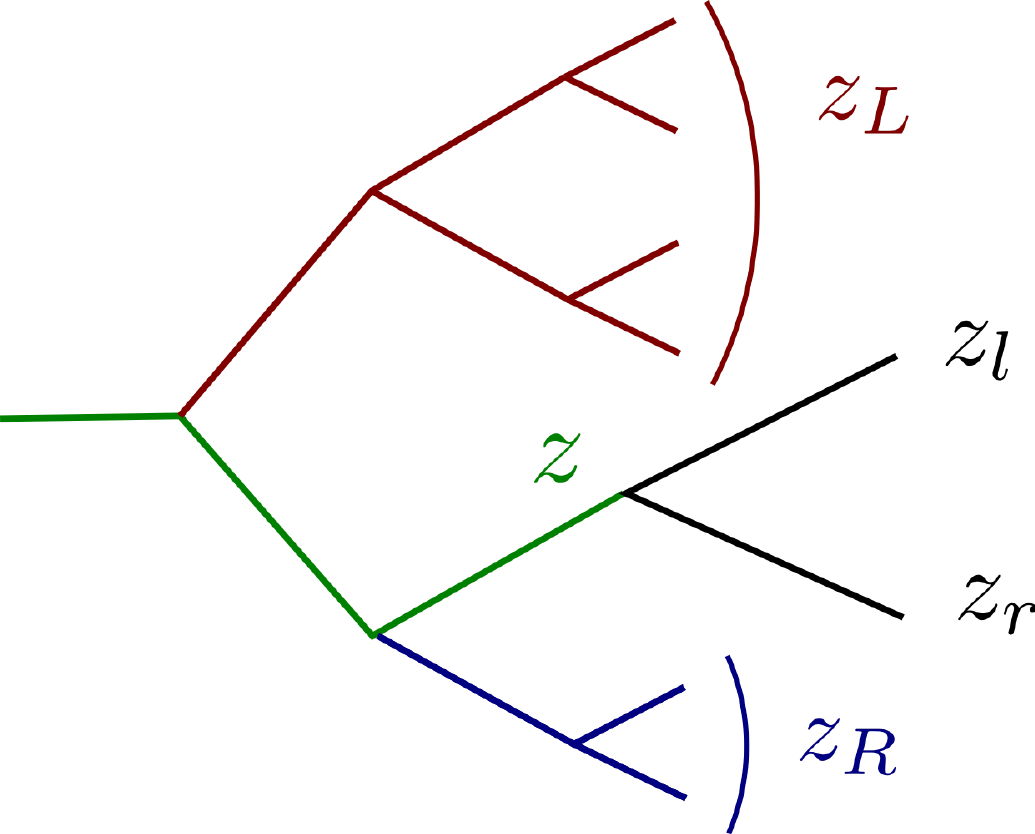}%
\caption{Parton shower picture for the broadening axis. Here all momentum fractions are relative to the parton initiating the jet. When the branch along which the axis lies (green) splits, the energy fractions $z_L$ and $z_R$ of the left (red) and right (blue) affect which daughter (black) gets  the axis.}
\label{fig:planarTree}
\end{figure}

For simplicity, we first consider the case where all the particles lie in a plane. In this case $\hat n$ is parametrized by an angle $\varphi$,
\begin{align} \label{eq:planarBroad}
	b(\varphi) = \sum_i z_i |\vartheta_i - \varphi|
\,.\end{align}
Since this function is piecewise linear, its minimum coincides with the the direction of one of the particles, $\varphi = \vartheta_i$. In the parton shower picture, we want to track the axis along the showering tree. To determine for a given splitting which of the two daughters takes control of the axis, requires comparing
\begin{align} \label{eq:leftVSright}
	b_l &= z_r(\vartheta_l-\vartheta_r) + \sum_{i \in L} z_i(\vartheta_i - \vartheta_l) +
	\sum_{i\in R}z_i( \vartheta_l - \vartheta_i) 
	\,, \nn \\ 
	b_r &= z_l(\vartheta_l-\vartheta_r) + \sum_{i \in L} z_i(\vartheta_i - \vartheta_r) + 
	\sum_{i\in R}z_i(\vartheta_r - \vartheta_i)
\,.\end{align}
Here $L$ and $R$ identify the subset of particles to the left and to the right of the splitting, see \fig{planarTree}. Subtracting the two lines in \eq{leftVSright} from one another gives
\begin{align} \label{eq:leftWins}
	b_l < b_r\quad \Leftrightarrow\quad z_l + z_L > z_r + z_R
\,,\end{align}
where $z_L$ and $z_R$ are the energy fractions of $L$ and $R$. In contrast to the WTA axis, it is thus not sufficient to compare $z_l$ and $z_r$, as the other branches still enter in \eq{leftWins}. It is still possible to determine the broadening axis with a recursive procedure, as long as one also keeps track of the total energy on the left/right of the axis. The algorithm reads:
\begin{enumerate}
        \item Start at the root of tree with initial condition $(z_L,z,z_R) = (0,1,0)$, where $z$ denotes the momentum fraction of the branch that tracks the axis.
        \item For the splitting $z = z_l + z_r$:
        \begin{itemize}
	\item If $z_L + z_l > \frac{1}{2}$, axis is along left daughter and $(z_L,z,z_R) \rightarrow(z_L,z_l,z_R+z_r)$.
	\item
	Otherwise, broadening axis is along right daughter and $(z_L,z,z_R) \rightarrow (z_L+z_l,z_r,z_R)$.
	\end{itemize}
        \item Repeat step 2 until there are no further splittings.
\end{enumerate}
This is thus described by a DGLAP evolution with two variables (one of the three can be eliminated, since $z_L + z + z_R =1$).

\begin{figure}[t]
\centering
\includegraphics[width=0.4\textwidth]{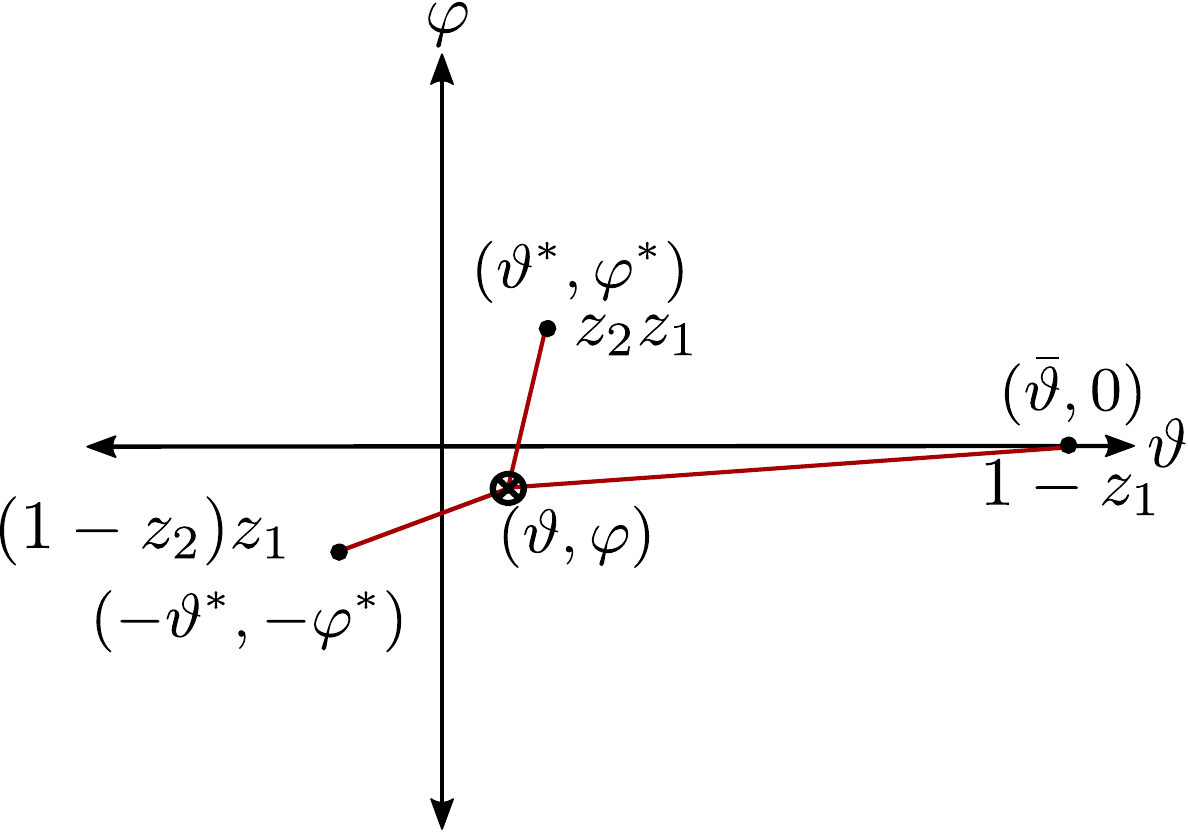}%
\caption{The three-particle configuration consists of a pair of particles and a third particle that is far away, due to the strong angular ordering. We also show the position of a test broadening axis, and indicate the angular distances (with red lines) that determine the broadening. The origin of the coordinate system is chosen to lie between the two nearby particles and the $\vartheta$ axis is chosen such that the third particle lies on it. }
\label{fig:BA3}
\end{figure}

We now move on to the non-planar case and consider the simplest non-trivial configuration of three particles, arising from two splittings. We will show that even in the strongly-ordered angular limit, the position of the broadening axis generically depends sensitively on the energy fraction of the initial splitting. This is in contrast to the WTA axis, where as long as the initial splitting is not with exactly balancing energy fractions, the axis will be stationary for any small perturbations of the initial energy fraction.\footnote{This is to say, the WTA axis will follow the branch with energy fraction greater than a half. We can change this energy fraction by a small amount, indeed, by any amount such that we maintain that the energy fraction of the initial splitting is still greater than a half, and the WTA axis will remain within the branch. If we leave all other \emph{relative} energy fractions further down the branch unchanged, it will even assume the same position.} Thus, the broadening axis does not behave in a Markovian manner with respect to a history in the strongly ordered angular limit. This is not the case for the WTA axis, whose position in the strongly ordered angular limit only depends on the branching that is currently occurring in the history. Without the Markovian condition, we do not expect the transverse momentum with respect to the broadening axis to have a simple leading logarithmic resummation.

After the first splitting, the broadening axis simply is along the particle with energy fraction $z_1>\frac{1}{2}$. We consider the case that this splits into a pair of particles with energy fractions $z_2z_1$ and $(1-z_2)z_1$, and choose the coordinate system in \fig{BA3}.
Using that for narrow jets the angular-separation measure of two particles is flat,
$\mathrm{d}\Omega \approx \mathrm{d}\vartheta^2 + \mathrm{d}\varphi^2$, the broadening for the axis along $(\vartheta, \varphi)$ is
\begin{align}\label{eq:broad3particles}
	b(\vartheta,\varphi) &= z_2z_1\sqrt{(\vartheta-\vartheta^*)^2 + (\varphi-\varphi^*)^2}+
	(1-z_2)z_1\sqrt{(\vartheta+\vartheta^*)^2 + (\varphi+\varphi^*)^2}
	\nn \\ & \quad
	+(1-z_1)\sqrt{(\vartheta-\bar\vartheta)^2+\varphi^2}.
\end{align}
We now search for a local minimum by taking derivatives, exploiting strong angular ordering $\bar \vartheta \to \infty$, and using 
\begin{align} \label{eq:angularRangePhi}
\vartheta^*, \varphi^* > 0
\,, \qquad
-\varphi^*< \varphi<\varphi^*
 \,, \qquad
-\vartheta^*< \vartheta < \bar\vartheta
\,.\end{align}
The last two conditions follow because the broadening axis lies within the convex hull of the three particles.

\begin{figure}[t]
\centering
\includegraphics[width=0.75\textwidth]{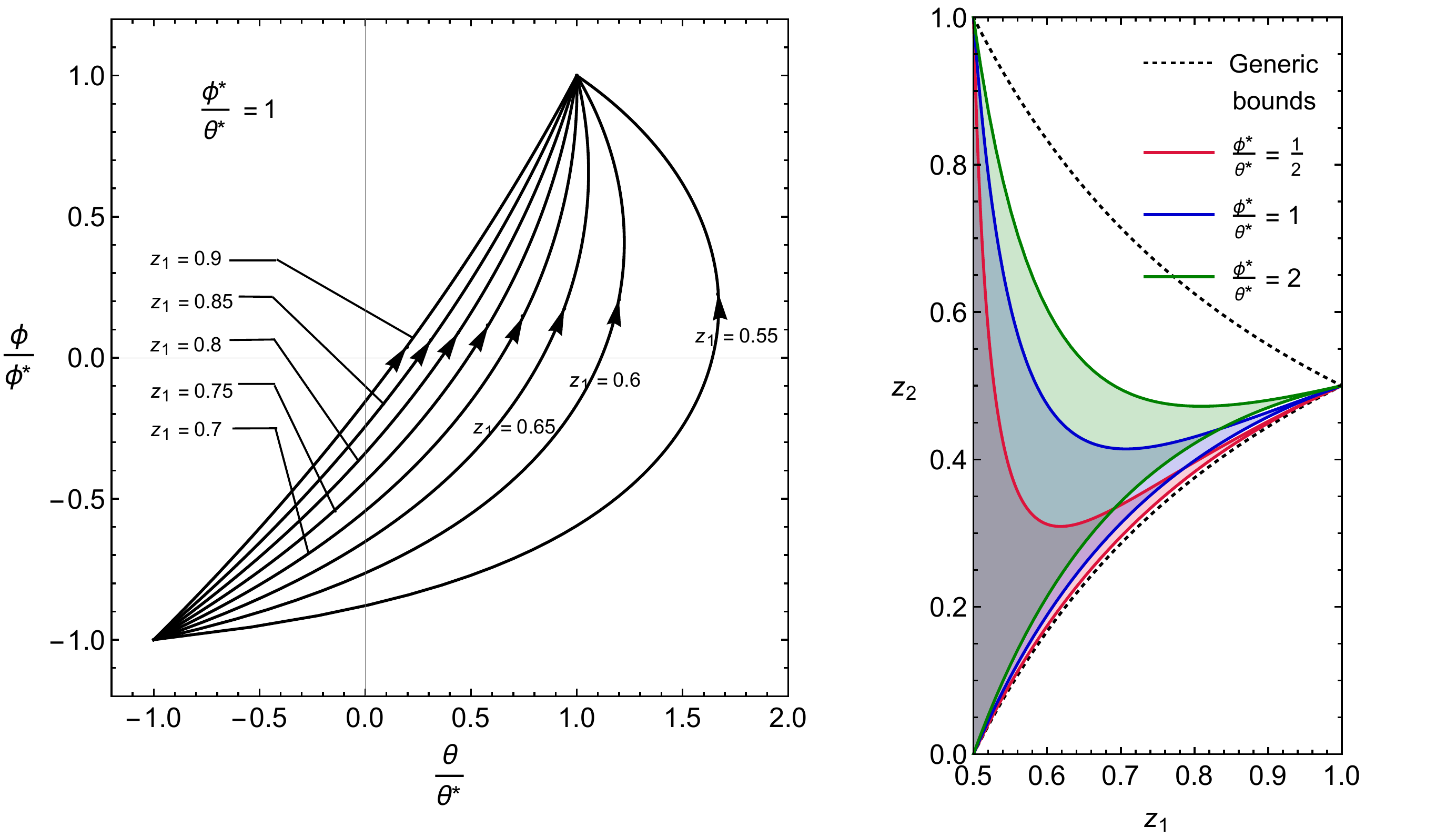}%
\caption{Left panel: The position of the broadening axis for $\varphi^*/\vartheta^* = 1$ is mapped out. The curves correspond to different $z_1$ values, showing that the Markovian condition is violated, and run over the allowed $z_2$ range.
Right panel: The region in $(z_1,z_2)$ where the broadening axis does \emph{not} lie on a particle. The different shades correspond to different values of $\varphi^*/\vartheta^*$. The generic bounds correspond to solutions of \eqref{eq:broad_bounds} without taking into account the additional constraints from \eq{angularRangePhi}.}
\label{fig:BAregions}
\end{figure}

It's convenient to switch to the variables
\begin{align} \label{eq:defEtaXi}
	\eta \equiv \frac{\vartheta-\vartheta^*}{\varphi-\varphi^*}
	\,, \qquad
	\xi \equiv \frac{\vartheta+\vartheta^*}{\varphi+\varphi^*}
\,,\end{align}
for which the condition of a local minimum takes the following form
\begin{align} \label{eq:broad3A}
	\xi - \eta &= 
         \frac{1-z_1}{z_2\,z_1}\, \sqrt{1+\eta^2}
	\,,\qquad
	z_2^2 (1+\xi^2) = (1-z_2)^2 (1+\eta^2)
\,.\end{align}
It has a solution for
\begin{align} \label{eq:broad_bounds}
     \frac{2z_1-1}{2z_1} < z_2 < \frac{1}{2z_1}
\,,\end{align}
which is given by
\begin{align} \label{eq:broad_sol}
     \eta &= \frac{-1+2z_1-2z_1^2 z_2}{\sqrt{-1\!+\!4z_1\!-\!4z_1^2\!-\!4z_1^2z_2\!+\!8z_1^3z_2\!+\!4z_1^2z_2^2\!-\!8z_1^3z_2^2}}
  \,,  \qquad 
   \xi = \frac{\sqrt{1\!+\!\eta^2\!-\!2z_2\!-\!2\eta^2 z_2\!+\!\eta^2 z_2^2}}{z_2}
\,.\end{align}
As in the planar case, the position of the broadening axis depends not just on two daughters of the splitting, but also the other particle (through $z_1$). However, unlike the planar case, the broadening axis does not have to lie on a particle, though it will do so for values of $z_2$ outside the bounds in \eq{broad_bounds}. Indeed, these boundaries exactly correspond to the condition that the momentum fractions of all of the partons are less than half, since the broadening axis will be along a parton if its momentum fraction is larger than half.

The picture is still slightly more complicated, because we did not yet impose \eq{angularRangePhi} on our solution, which shrinks the solution space. (In particular it should vanish in the planar limit.) Since the analytic expressions are rather complicated, we illustrate the effect in \fig{BAregions}. In the right panel, we show how the solution space shrinks depending on the ratio $\frac{\varphi^*}{\vartheta^*}$. In the left panel, we show the position of the broadening axis for $\frac{\varphi^*}{\vartheta^*}$ = 1, with different paths corresponding to different values of $z_1$. This dependence on $z_1$ explicitly shows that it violates the Markovian condition.

In conclusion, we have shown that at leading-logarithmic order (i.e.~strong angular ordering) the broadening axis generically does not lie on a particle, and depends on particles other than the daughters of the splitting under consideration. It is therefore clear that there is no simple DGLAP evolution that describes this parton shower picture, though it can of course be calculated by simulating the \emph{full} shower.

\subsection{Jet shape}
\label{sec:jetshape}

The jet shape is the average fraction of the jet energy at a specific angle $\theta$ with the axis. The corresponding cross section can be obtained from \eq{match_si}, by expressing $\vec k_\perp$ in terms of $\theta$ using \eq{angle}, summing over hadron species $h$ and averaging over $z_h$. Explicitly for $Q \gg E_J R \gg k \gg \lqcd$, we combine eqs.~\eqref{eq:match_si}, \eqref{eq:match_G} and \eqref{eq:match_D} to obtain 
\begin{align} \label{eq:jetshape}
\frac{\df \langle z\rangle}{\df E_J\, \df \theta}
  &= \frac{2\pi \theta E_J^2}{\si} \sum_h \int\! \df z_h\, z_h\,
  \frac{\df \si_h}{\df E_J\, \df^2 \vec k_\perp\, \df z_h}
  \nn \\
  &=  \frac{2\pi \theta E_J^2}{\si} \sum_{i,j,k} \int\! \frac{\df x}{x}\, H_i\Big(\frac{E_J}{x}, \mu\Big) 
    \int\! \df y\, y\,B_{ij}(x, E_J R, y,\mu) 
    \int\! \df z\, z\, C_{jk}(\vec k_\perp, z,\mu)
  \nn \\ & \quad \times
  \sum_h \int\! \df z_h\, z_h\, d_{k \to h}(z_h,\mu)   
.\end{align}
The overall factor is the Jacobian due to switching from $\vec k_\perp$ to $\theta$, and the full cross section $\si$ that we normalize to. The dependence on fragmentation functions drops out because of the momentum sum rule
\begin{align} \label{eq:sumrule}
  \sum_h \int\! \df z_h\, z_h\, d_{k \to h}(z_h,\mu)  = 1
\,.\end{align}

The jet shape can also be defined on subsets $S$ of particles, such as charged particles, restricting the sum over $h$ in \eq{jetshape}. In that case we cannot use \eq{sumrule} to eliminate the fragmentation functions completely. However, the required nonperturbative input is rather limited, as we need one nonperturbative number for each parton flavor, which describes the average energy of a parton that goes into particles in the subset $S$.

\subsection{Threshold effects}
\label{sec:threshold}

Because the jet shape turns off sharply at $\theta = R$, the endpoint in terms of $|k_\perp| = E_J R = x_J QR/2$ is rather sensitive to the $x_J$ distribution. In particular, in the vicinity of this endpoint, threshold logarithms of $1-x_J$ need to be resummed (given that the distribution is peaked around $x_J=1$, this resummation is justified in general). We will include these double logarithms at LL accuracy, to capture the dominant behavior of this effect. 

\begin{figure}[t]
\centering
\includegraphics[width=0.4\textwidth]{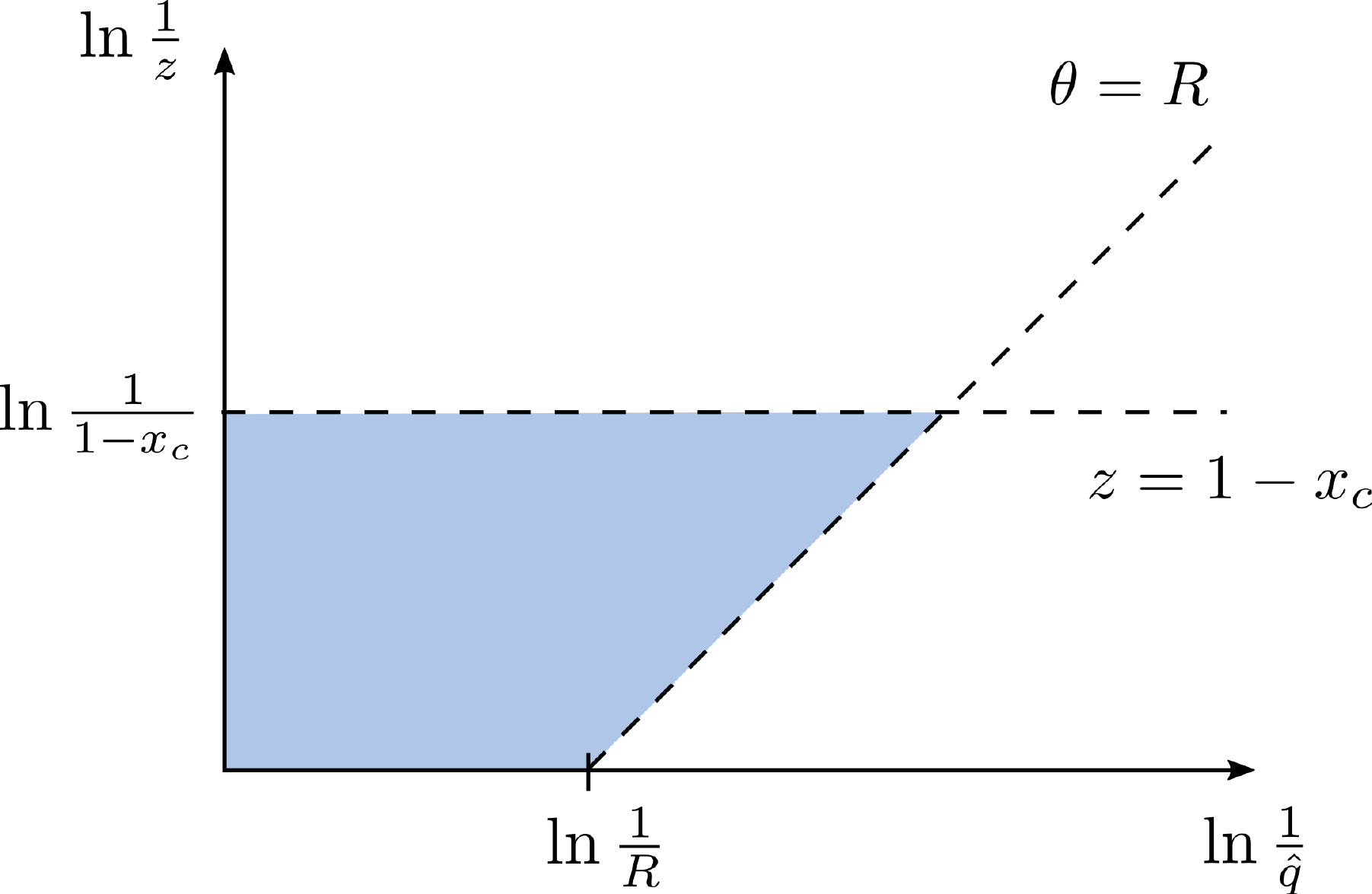}%
\caption{Requiring $x \geq x_c$ prohibits emissions in the (blue) shaded region of phase-space, in terms of the momentum fraction $z$ and transverse momentum $\hat q_\perp$ in \eq{qperp} of the emission.}
\label{fig:sudakov_ps}
\end{figure}

To argue their form, we use again a parton shower picture, now strongly-ordered in angle \emph{and} energy.
Requiring $x_J \geq x_c$, prohibits emissions with angle $\theta$ and momentum fraction $z$
\begin{align} \label{eq:thr_cut}
 \theta > R \quad
 \text{and} \quad
 z > 1-x_c
\,.\end{align}
For technical reasons\footnote{In the collinear and soft approximation, \eq{coll_approx} becomes $\df z/z^{1+2\eps}\, \df \theta/\theta^{1+2\eps}$, using dimensional regularization. Now $\int_R^\infty\!\df \theta/\theta^{1+2\eps} = 1/(2\eps) - \ln R + \ord{\eps}$, and the $1/(2\eps)$ gives a finite contribution when combined with the $2\eps \ln z$ from $z^{1+2\eps}$, so we cannot set $\eps=0$ at the beginning of the calculation. However, this \emph{can} be done when using $\hat q_\perp$, since then we have $\df z/z\, \df \hat q_\perp/\hat q_\perp^{1+2\eps}$.}, it is convenient to use the (dimensionless) transverse momentum
\begin{align} \label{eq:qperp}
 \hat q_\perp = z (1-z) \theta
\,, \end{align} 
rather than the angle $\theta$ itself. The phase-space boundaries corresponding to \eq{thr_cut} in terms of the coordinates $\ln (1/z)$ and $\ln (1/\hat q_\perp)$ are shown in \fig{sudakov_ps}. At LL accuracy, the emission probability is uniform in terms of these coordinates, so we can simply calculate the shaded area. Also, emissions can be treated as independent at this order, leading to the following Sudakov factor, 
\begin{align} \label{eq:sudakov}
  \int_{x_c}^1\, \df x_J\, \frac{\df \si^\thr_i}{\df x_J} = \si_{i} \exp \bigg\{-\frac{2 \al_s C_i}{\pi} \Big[ \ln(1-x_c) \ln R + \frac12 \ln(1-x_c)^2 \Big]\bigg\}
\,,\end{align}
where we split $\si = \si_q + \si_g$, and $C_i = C_F$ for quark jets and $C_A$ for gluon jets.

Even if we are only interested in the cross section integrated over the jet energy, we shouldn't implement threshold corrections after this integral, since this neglects correlations between the jet energy $E_J$ and the transverse momentum $\vec{k}_\perp$. Instead, we apply the Sudakov factor in \eq{sudakov} to the $x_J$ spectrum, 
\begin{align} \label{eq:sudakovDifferential}
  \frac{\df \si_h}{\df x_J\, \df^2 \vec k_\perp\, \df z_h} &= -\frac{Q}{2} \sum_i \frac{\df}{\df x'}\bigg[ \int_{x'}^1\! \df x'' \int_{x''}^1\frac{\df x}{x}\, \tilde{H}_i\Big(\frac{x''Q}{2x},\mu\Big)\tilde{\cG}_i(x,x_J Q R/2,\vec{k}_\perp,z_h,\mu)
	\nn \\ & \quad \times
	\exp\Big\{-\frac{\al_s C_i}{\pi}\ln^2(1-x')\Big\}\bigg]_{x' = x_J}\,.
\end{align}
Here we took the cumulative of \eq{match_si}, then included the threshold Sudakov in \eq{sudakov}, and took the derivative. The main point is that the $E_J$ that appears in $\cG$ corresponds to the measured $x_J$ (and not $x$, $x'$ or $x''$). A direct comparison with our NLO ingredients reveals that the $\ln R$ term in \eq{sudakov} is already accounted for by small-$R$ resummation, and we therefore omitted it to avoid double counting. Similarly, we subtracted the overlap between the exponentiated threshold logarithms and the NLO expressions for $\cG$ and $H$, as indicated by the tilde.
Although the implementation of \eq{sudakovDifferential} directly follows from \eq{sudakov}, it is not very efficient. We therefore use the following simpler prescription, 
\begin{align} \label{eq:easySudakov}
  \frac{\df \si_h}{\df x_J\, \df^2 \vec k_\perp\, \df z_h} &= -\frac{Q}{2} \sum_i \int_{x_J}^1\! \frac{\df x''}{x''} \int_{x''}^1\frac{\df x}{x}\, \tilde{H}_i\Big(\frac{x''Q}{2x},\mu\Big)\tilde{\cG}_i(x,x_J Q R/2,\vec{k}_\perp,z_h,\mu)
	\nn \\ & \quad \times
	\frac{\df}{\df x'} \bigg[\exp\Big\{-\frac{\al_s C_i}{\pi}\ln^2(1-x')\Big\}\bigg]_{x' = x_J/x''}\,,
\end{align}
which holds to the same accuracy.

\subsection{Numerical Implementation}
\label{sec:numerical}

We now present in some detail our implementation, starting with how we solved the evolution equations numerically. We then show how we combine the factorization theorems for $k \ll E_J R$ and $k \sim E_J R$, and conclude with a discussion of our scale choice and uncertainty estimates. The one-loop expressions for $H$ are well known~\cite{Altarelli:1979k,Furmanski:1981cw,Nason:1993xx,Berger:1995fm}. The coefficients $B$ in \eq{match_G}, $C$ in \eq{match_D} and $J$ in \eq{match_G_bis} were calculated at one-loop order in ref.~\cite{Neill:2016vbi}, where in our implementation the relation $p_T R \leftrightarrow 2E_J\tan\frac{R}{2}$ is used to convert the $pp$ jet definition (that uses azimuthal angles and pseudo-rapidity) to our $e^+e^-$ case (that uses angles). Keeping the tangent (with respect to the angle $R$) retains some subleading corrections in the small-$R$ expansion underpinning \eq{match_si}, improving the behavior of the cross section in the vicinity of the $\theta \sim R$ endpoint. Similarly, we capture more (but not all) subleading corrections by using $\theta(p_T R - k) \leftrightarrow \theta(E_J\sin{R} - k)$ in the jet functions $J$.

The starting point for all the numerical results shown in \sec{results} is \eq{match_si}. We work with cumulative distribution in $\vec{k}_\perp$,
\begin{align} \label{eq:intKperp}
  \int_{k< k_c}\!\!\! \df^2 \vec{k}_\perp \ \frac{\df \sigma_h}{\df^2\vec{k}_\perp\, \df E_J\, \df z_h}
  = \pi \int_0^{k_c}\! \df k^2 \, \frac{\df \sigma_h}{\df^2\vec{k}_\perp\, \df E_J\, \df z_h}\, ,
\end{align}
differentiating the result at the end of the computation. This does not complicate the evolution, because the anomalous dimensions do not involve $\vec k_\perp$. In fact, it is necessary to choose the scales of the evolution in terms of $k_c$, because $C^{(0)}_{ij}(\vec{k}_\perp) \propto \delta^2(\vec{k}_{\perp})$ would cause the distribution to vanish unless $\vec k_\perp = 0$. By contrast, $C^{(0)}_{ij}(k_c) \propto \theta(k_c)$ as function of $k_c$.

The resummation is implemented in the form presented in~\eq{RGE}: we start from the fragmentation functions $d$ at some initial scale $\mu_d$ and we evolve them to $\mu_C \sim k_c$ where we match onto the TMD fragmentation function $D$. We note that the convolution variable of the evolution is the energy fraction of the hadron $z$. We then evolve the TMD fragmentation function using the modified DGLAP to $\mu_B \sim E_J R$, with the convolution variable still effectively being the energy fraction of the hadron. At the $\mu_B$ scale, we match onto the fragmenting jet function $\cG$. Finally we evolve using standard DGLAP up to $\mu_H \sim Q$, with the convolution variable now being the energy fraction of the hard parton which initiates the jet, denoted by $x$. At this point the corrections from the hard function are included.\footnote{We stress that due to the matrix nature of the factorization formulae the various evolution/matching steps do not commute.} We repeat the evolution separately for different values of the jet energy fraction $E_J$ and finally integrate over it. Each term in the NLO corrections $H^\one,C^\one,B^\one,J^\one$ only has a nontrivial dependence on $x$ or $z$ but not both, so these evolutions factorize and are carried out separately.

All RGEs are solved using the classic Runge-Kutta method in the evolution basis (singlet/nonsinglet decomposition) and account for heavy quark thresholds. However, we adopt different strategies depending on the observable. The jet shape in \eq{jetshape} is the first moment in $z_h$, so in this case we find it natural to perform the evolution in Mellin space, where it becomes multiplicative. The one-loop anomalous dimensions for (modified) DGLAP evolution in Mellin space were given in ref. \cite{Neill:2016vbi}\footnote{We found a typo in eq. (C.3) there. The correct expression for the $(gg)$ modified anomalous dimension is \\ ${\gamma'}_{gg}^\one (N,\mu) = \gamma^\one_{gg}(N,\mu) - \tfrac{\al_s(\mu) C_A}{\pi}\big[-2H_{1/2}(N\!+\!1)+2\ln 2 + 2^{-N-2}\frac{5N^3+33N^2+68N+48}{N(N+1)(N+2)(N+3)}\big].$}. When inclusive over all hadrons, we can use the sum rule in \eq{sumrule} to remove input from fragmentation functions, while for charged pions we take the first moment of existing parameterizations provided by the latest DSS~\cite{deFlorian:2014xna} and HKNS~\cite{Hirai:2007cx} (global) fits. Even if we perform LL, the presence of NLO fixed-order ingredients justifies the usage of their NLO sets, where more recent parametrizations are available. As a different observable, we consider the cross section differential in the hadron energy fraction $z_h$, varying cuts on the transverse momentum $k_c$. In this case we carry out the evolution directly in $z$ space, performing the Mellin convolutions on a 75-step linear grid.

Small-$R$ resummation in $e^+ e^-$ collisions poses the additional issue of evolving distributions in the convolution variable, such as $\delta(1-x)$, rather than functions (for the convolutions in $z$ there is no problem, as such distributions are smeared by their convolution with phenomenological fragmentation functions). We solve this issue by taking the zero truncated moment of the distribution and exploiting that such a truncated moment itself satisfies a DGLAP evolution equation with modified splitting functions~\cite{Kotlorz:2006dj}. This is simply due to the rearrangement 
\begin{align} \label{eq:truncated_mom}
 \int_{x_0}^1\! \df x \int_x^1\! \frac{\df x'}{x'}\, f(x') g\Big(\frac{x}{x'}\Big) = 
 \int_{x_0}^1\! \frac{\df x'}{x'}\, \frac{x_0}{x'}\, f\Big(\frac{x_0}{x'}\Big)
  \int_{x'}^1\! \df x\, g(x)
\,,\end{align}
where in our case $f$ is the splitting function. To get the evolved
spectrum we then differentiate the evolved truncated moment. To
validate this technique, we checked its accuracy against evolution in
Mellin space for the jet shape differential in angle and for additional test observables. We found good agreement on the normalized results we show, well within our uncertainty bands. A non-negligible difference persists in absolute normalization, caused by the large sensitivity of the method on the $x=1$ endpoint, where our distributions are peaked. To mitigate this effect, we use a 60-point exponential grid (that becomes finer as $x \to 1$), and we only show normalized results. Finally, we resum threshold logarithms of $1-x_J$ at LL accuracy, as discussed in \sec{threshold}. Using equation \eq{easySudakov} this only requires an additional convolution step.

We now describe how to we extend predictions from $k_c \ll E_J R$ to $k_c \lesssim E_J R$, referring for definiteness to the jet shape. For large $k_c$ one has to include the power corrections in \eq{jetshape}, by using the coefficients on the left hand side \eq{match_J}. Since $\mu_C \sim \mu_B$, we turn off resummation between the two scales. In order to transition between the two regimes in a continuous way, we schematically perform the matching
\begin{align}\label{eq:LLtoFOmatching}
\si &= H(Q) \otimes_x U(Q, E_J R) \otimes_x \Big[B(E_J R) \otimes_z U'(E_J R, k_c) \otimes_z C(k_c) \otimes_z d(k_c)
\nn \\ & \quad
 +  J(E_J R)\otimes_z d(E_J R) - B(E_J R) \otimes_z C(E_J R)\otimes_z d(E_J R)\Big]
\,,\end{align}
where $U$ ($U'$) are (modified) DGLAP evolution kernels between the indicated scales, and the arguments of the functions indicate the canonical scale at which they are evaluated. Interestingly, in the case of the jet shape differential in angle the second line vanishes for $\theta <R$, only cutting off the spectrum at $\theta=R$. We do not expect this to hold at higher order. 

We end this section discussing our choice of scales. Our standard configuration is
\begin{align} \label{eq:muH} 
   \mu_H = Q, \qquad \mu_B = 2E_J\tan (R/2), \qquad \mu_C = k_c\,,
\end{align}
which differs slightly from the canonical scales in \eq{canonical}. To avoid $\alpha_s(\mu_C)$ from hitting the Landau pole, we use the following profile for the scale $\mu_C$,
\begin{align}\label{eq:lowKprofile}
  \mu_C(k_c) = \frac{k_0}{2}\Big(1+\frac{k_c^2}{k_0^2}\Big) \qquad \text{if}\quad k_c \leq k_0\,,
\end{align}
where $k_0 = 1$ GeV. Of course in this region nonperturbative corrections to \eq{match_D} will be important. In principle we could also use a profile scale to gradually switch off resummation when approaching the region $k_c \sim E_J R$, see e.g.~refs.~\cite{Ligeti:2008ac,Abbate:2010xh}, but we find no clear reason to do so in our case.

Finally, we estimate the perturbative uncertainty by varying the scales by a factor 2. Specifically, we first only vary the profile scale $\mu_C$ around its central value, then vary $\mu_B$ and $\mu_C$ at the same time, and then vary all the three scales together. This probes the resummation uncertainties related to logarithms of $k_c/E_JR$, $R$ and fixed-order uncertainty, respectively. In addition, we vary the profile parameter $k_0$ up and down by a factor of 2. We take the final uncertainty to be the quadrature of the four cases.  We also investigated the dependence on the scale at which the coupling $\al_s(\mu)$ in threshold corrections in \eq{easySudakov} is evaluated, finding that this effect is negligibly small.

\section{Results}
\label{sec:results}

\begin{figure}[t]
\centering
\includegraphics[width=0.4\textwidth]{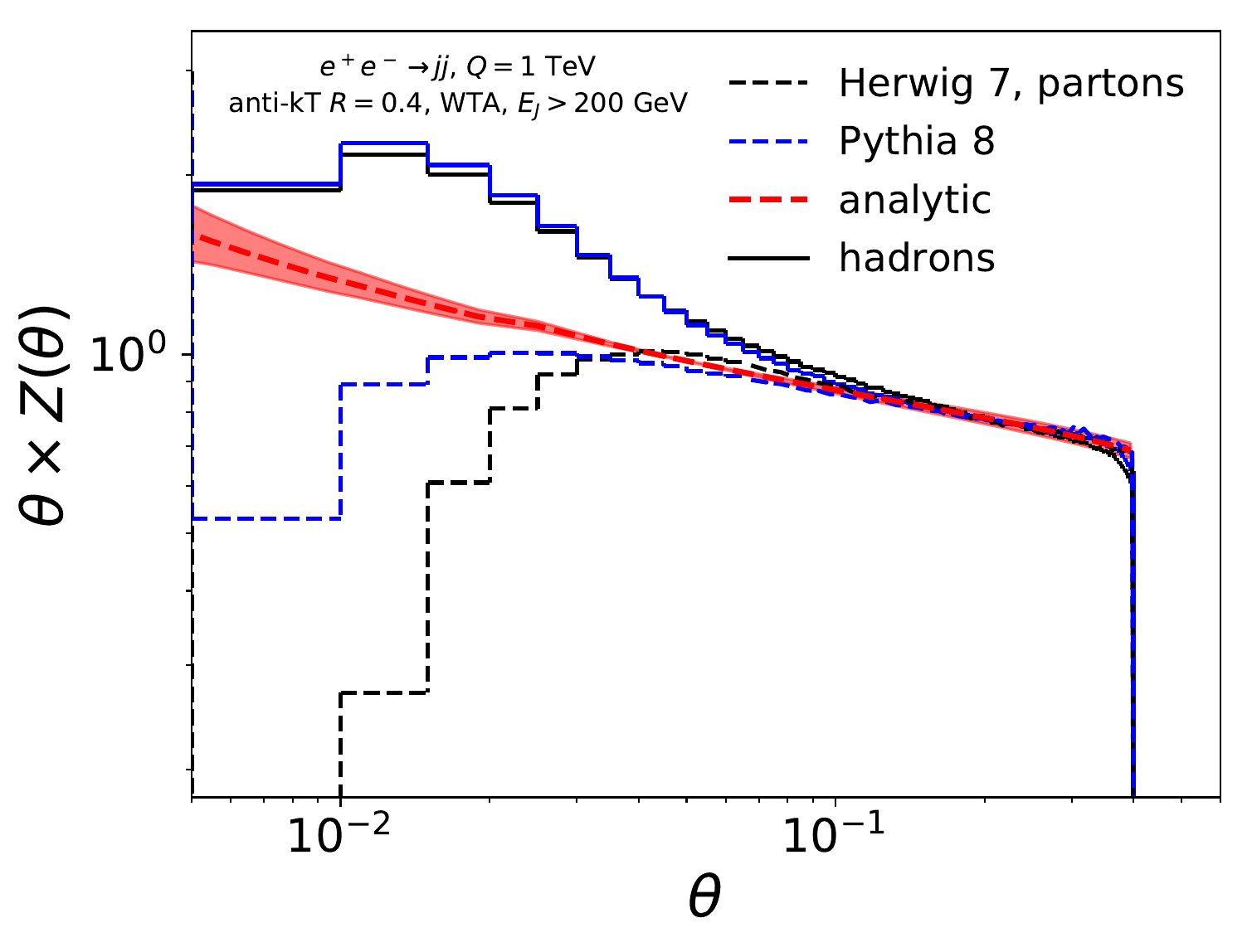} \qquad
\includegraphics[width=0.4\textwidth]{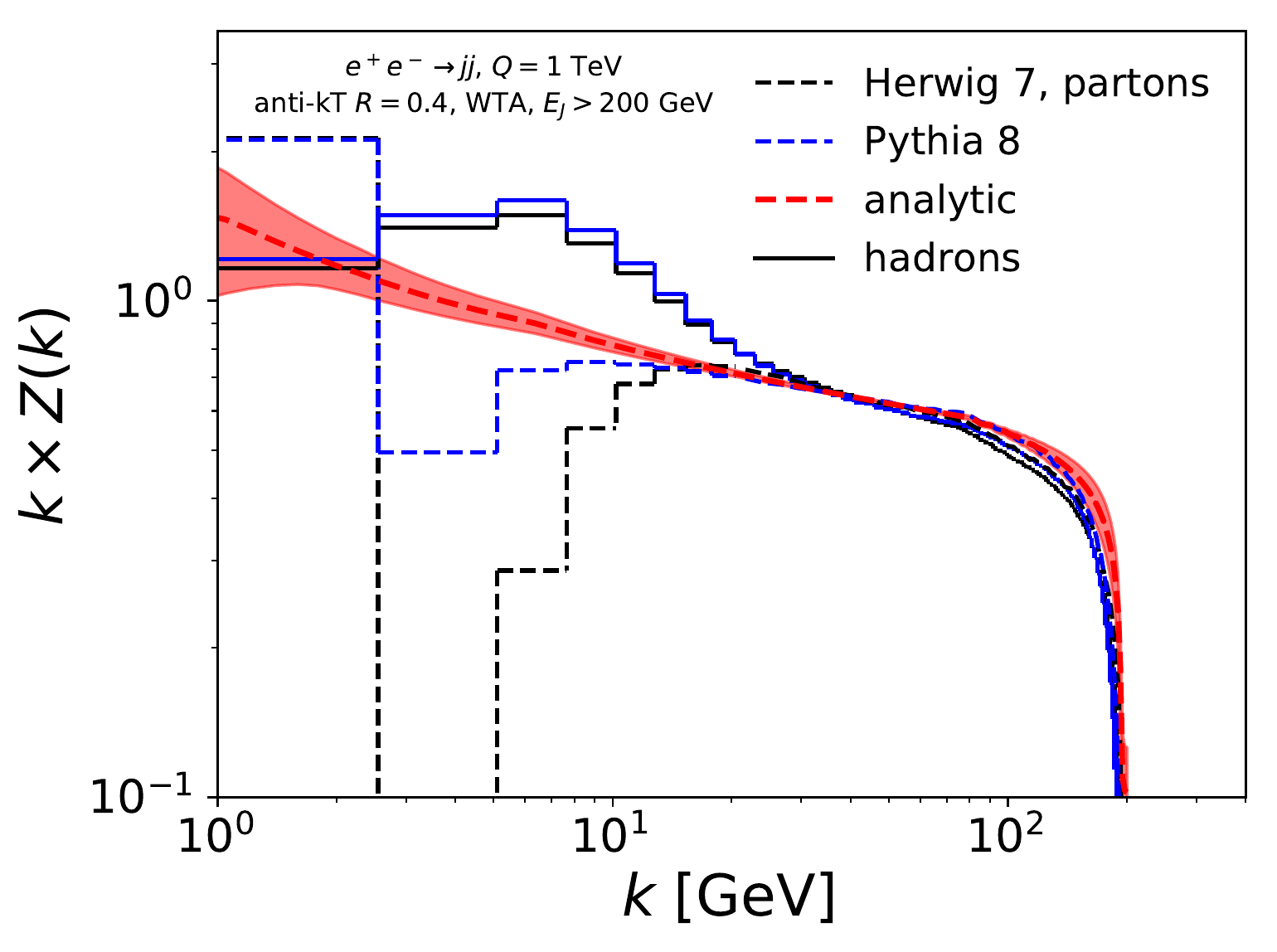} %
\caption{The jet shape as function of angle (left panel) and transverse momentum (right panel), predicted by \Herwig (black), \Pythia (blue) at parton (dashed) and hadron-level (solid), and obtained from our analytic calculations including perturbative uncertainties (red curve with band).}
\label{fig:zint_analytic}
\end{figure}

In this section we show results for the jet shape, differential in
angle or transverse momentum, and the fragmentation spectrum with a
cut on angle or transverse momentum. Our default setup is as follows:
$e^+ e^- \to$ jets at a center of mass energy of $Q = 1$ TeV. The jets
are identified using (the $e^+e^-$ version of) anti-$k_T$ with $R=0.4$
and the WTA recombination scheme, and jets are required to have jet
energy $E_J > 200$ GeV. We compare predictions from \Herwig 7.1.1 and \Pythia 8.226, through a \Rivet analysis~\cite{Buckley:2010ar}, to our analytic calculations using the framework in \sec{framework}. We choose the following normalization of the jet shape as our default,
\begin{align} \label{eq:Z_def}
Z(\theta) = 
\Big(\frac{\df \langle z \rangle}{\df
  \theta}\Big)\Big/\Big(\int_{\theta_{\rm min}}^{R}\! \df \theta\,\frac{\df \langle z \rangle}{\df \theta}\Big)
\,,\nn \\
Z(k) = 
\Big(\frac{\df \langle z \rangle}{\df k}\Big)\Big/\Big(\int_{k_{\rm
  min}}^{k_{\rm max}}\! \df k\,\frac{\df \langle z \rangle}{\df k}\Big)
\,.\end{align}
with $\theta_\mathrm{min} = 0.1$ and $k_{\rm min}=20$~GeV, $k_{\rm max}=100$~GeV. 

We start by showing in \fig{zint_analytic} results for the jet shape. The central region of the distribution follows an (approximate) power law, where deviations from a simple $1/\theta$ are due to the resummation of logarithms of $R/\theta$ and the running of the coupling constant. We have highlighted these deviations by plotting $\theta Z(\theta)$ rather than $Z(\theta)$. All predictions agree in this region, and ours have the added benefit of a theory uncertainty estimate. This power-law behavior extends to the edge of the jet for the angular distribution, but has a smooth turn off for the transverse momentum distribution (right panel) due to \eq{angle} and the jet energy distribution. In particular, reliable predictions near the endpoint require the resummation of threshold logarithms, discussed in \sec{threshold}, without which our predictions would disagree with \Herwig and \Pythia in this region.
Moving on to the small $k$ region, we note that the parton-level distribution is peaked near $k = 0$ because the WTA axis is always along a particle. The adjacent ``dead cone" is due to the shower cut off, and is filled with radiation by the hadronization model. This effect is normally not visible, because the position of the axis is smeared by soft radiation, suggesting that the winner-take-all axis is particularly useful for studying this nonperturbative physics. Note that \Herwig and \Pythia differ before hadronization, but agree well after including it. Our purely perturbative calculation is not valid in this region, and would require the inclusion of nonperturbative effects. The reason that nonperturbative effects already become important for $k \lesssim 10$ GeV, is due to our definition in \eq{k_def}. For example, a hadron with $z_h \sim 0.1$ and $k = 10$ GeV has a transverse momentum of $z_h k = 1$ GeV.

\begin{figure}[t]
\centering
\includegraphics[width=0.4\textwidth]{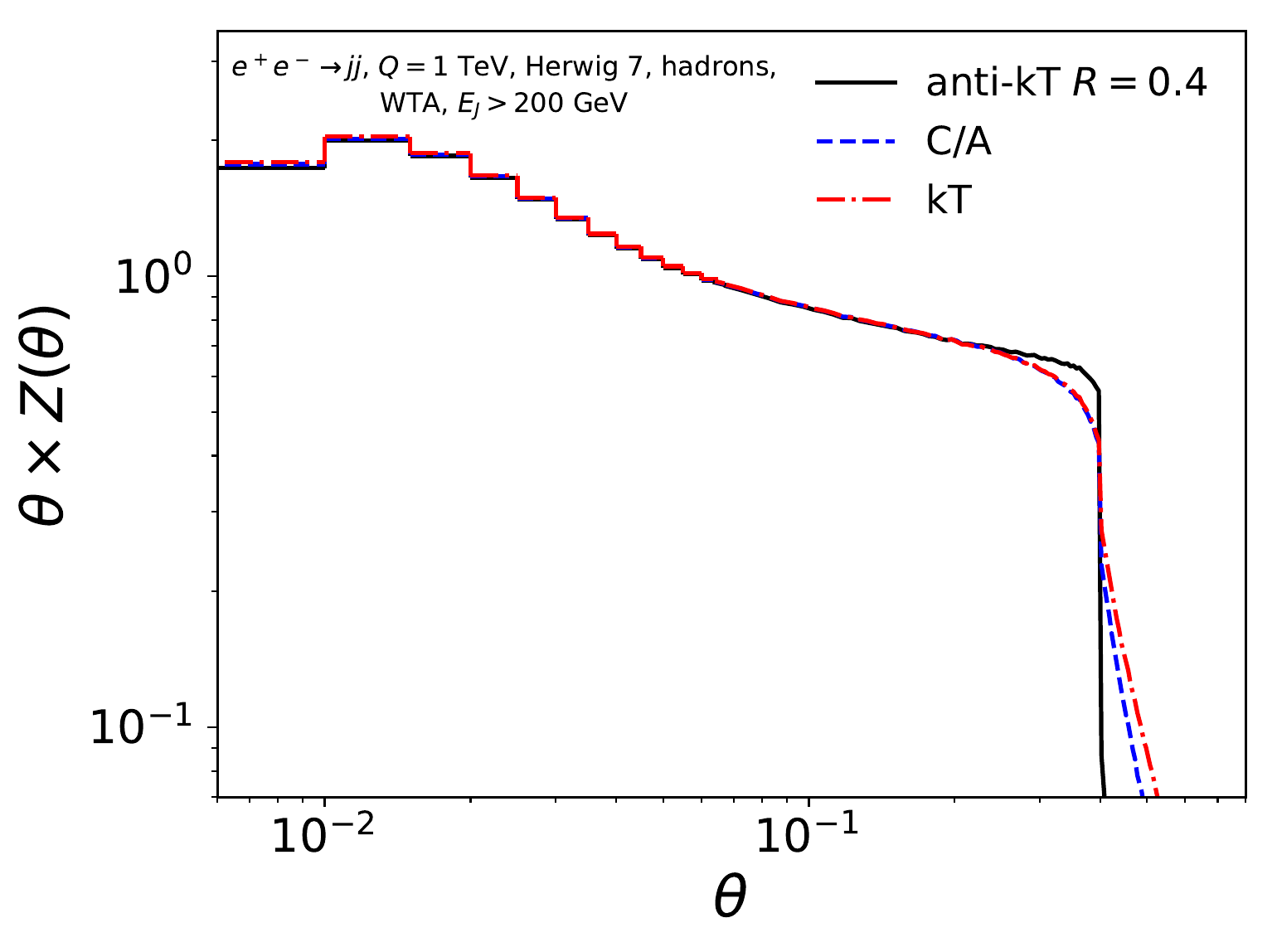}
\caption{The jet shape at hadron level as function of angle for the $e^+e^-$ anti-k$_T$ (black), Cambridge/Aachen (blue dashed) and $k_T$ algorithm (red dot-dashed). }
\label{fig:zint_jetalgs}
\end{figure}  

Next we investigate in \fig{zint_jetalgs} the dependence of the jet shape on the jet algorithm, comparing the ($e^+e^-$ version of) anti-$k_T$, Cambridge/Aachen and $k_T$ algorithms. There are only differences at the very edge of the jet, and they are rather small. Since the WTA axis is robust, these differences are due to particles at the edge of the jet being clustered into it or not. As expected, anti-$k_T$ has the sharpest edge and $k_T$ the softest edge. The differences between algorithms will become larger when there are many jets in an event, e.g.~when the cut on $E_J$ is loosened or $pp$ collisions are considered. 

\begin{figure}[t]
\centering
\includegraphics[width=0.4\textwidth]{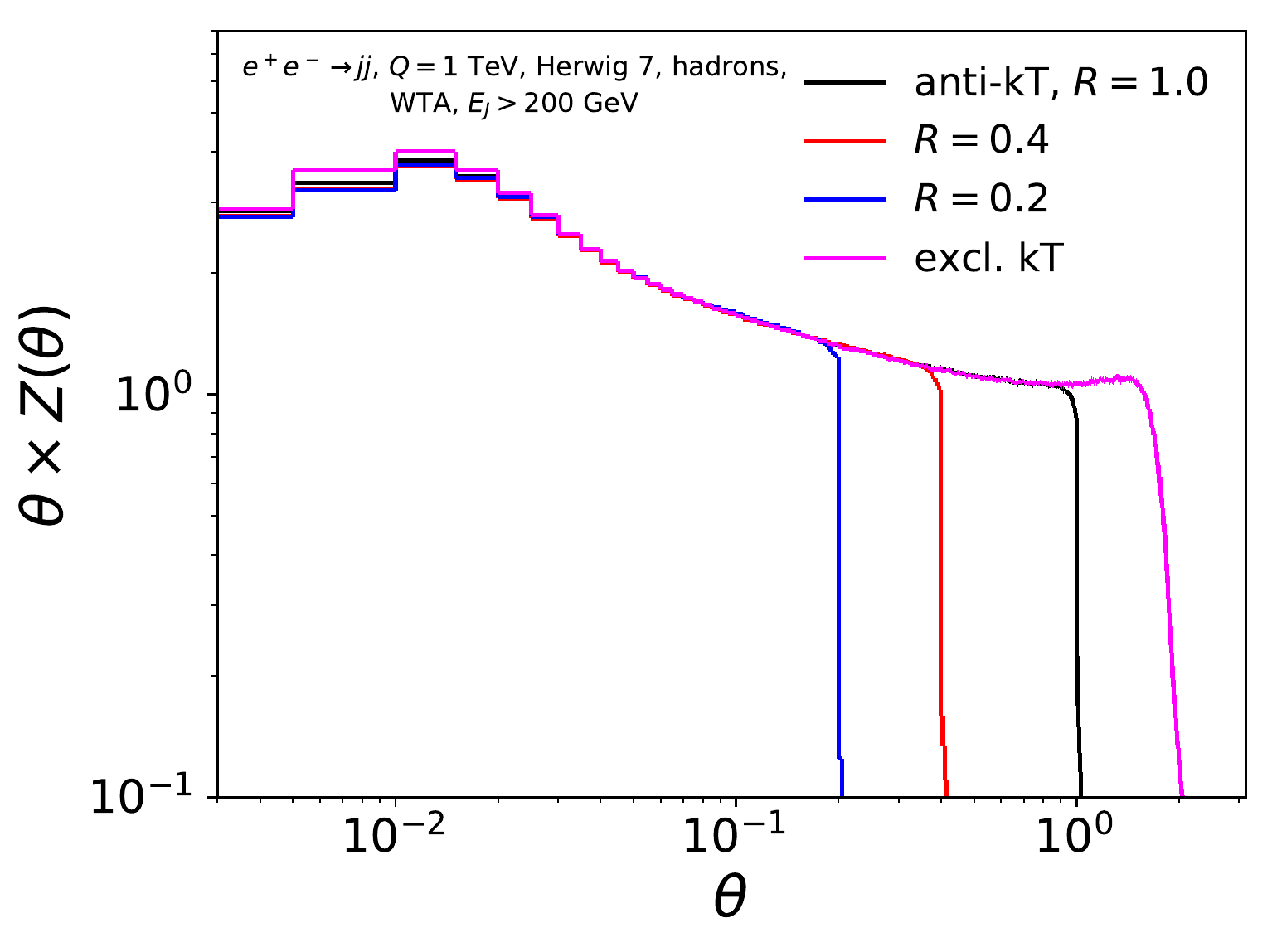} \qquad
\includegraphics[width=0.4\textwidth]{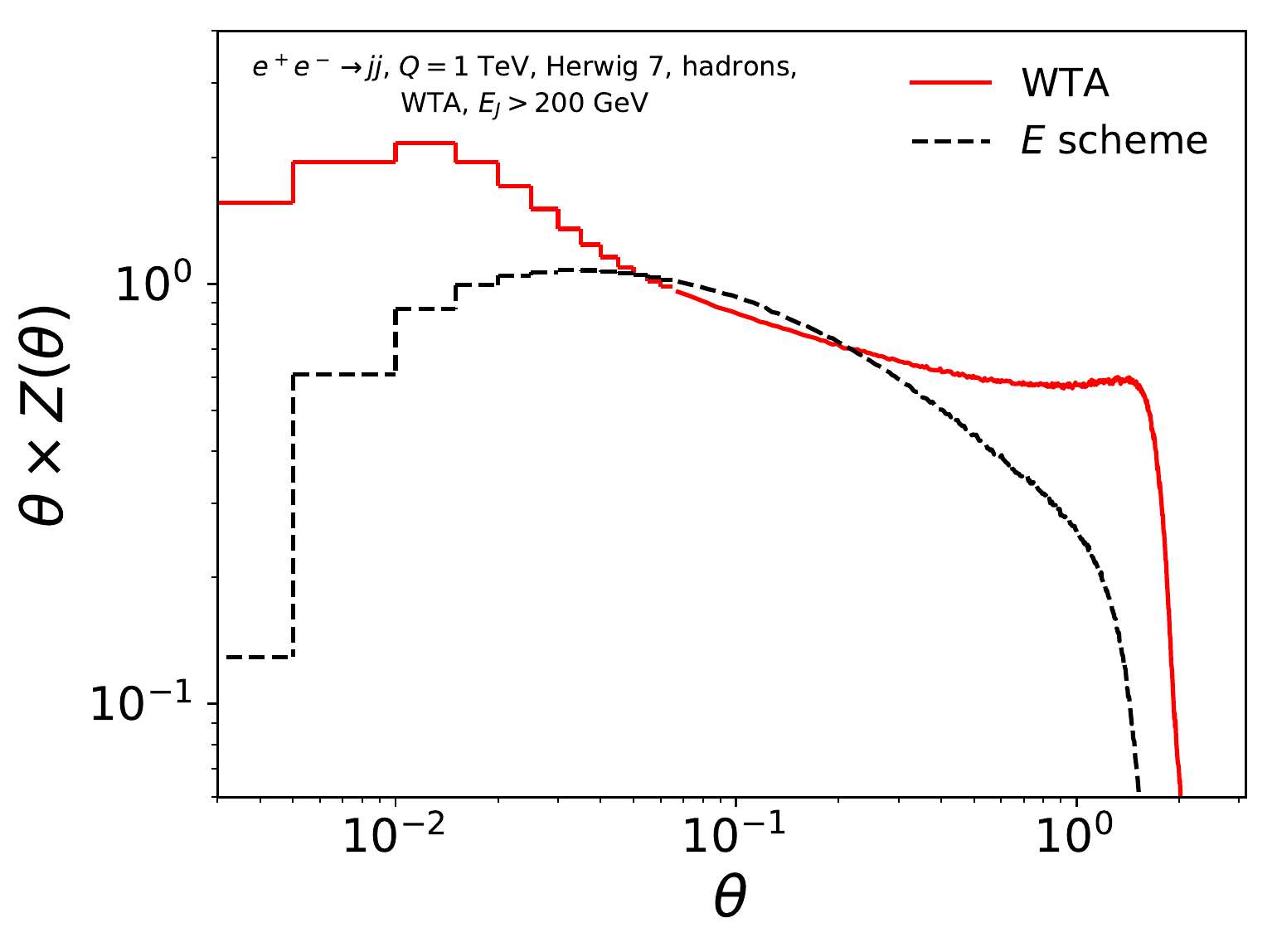}
\caption{Left panel: The jet shape as function of the angle for different values of the jet radius $R$, and for exclusive $k_T$. The curves are normalized in $\theta = [0.1, 0.2]$. Right panel: The jet shape for exclusive $k_T$ (two jets) as function of angle with the WTA axis (red) and standard jet axis (black dashed).}
\label{fig:zint_Rvariation}
\end{figure}

In the left panel of \fig{zint_Rvariation} we show the jet shape for
anti-$k_T$ with $R=0.2$, $0.4$ and $1$. Exclusive $k_T$ is also shown,
which clusters the whole event into two jets, and thus corresponds to
$R\sim \pi/2$. The WTA axis is the same, independent of $R$, which is why the distributions overlap. For larger values of $\theta$ the collinear approximation no longer holds, and the distributions even rises due to a Jacobian factor. Specifically, one  would expect a constant energy density from soft radiation, i.e.~$\df E/\df\Omega = \df E/(\df \phi \df \theta \sin \theta) \sim$ constant, implying that $\theta \df E/\df \theta \sim \theta/(\sin \theta)$ rises. This is not the case for the jet shape using the standard jet axis, as shown in the right panel of \fig{zint_Rvariation}, because the axis will reposition itself depending on all the radiation in the jet. This figure also clearly shows that the jet shape with standard jet axis exhibits Sudakov double logarithms instead of a power-law dependence on $\theta$.

\begin{figure}[t]
\centering
\includegraphics[width=0.317\textwidth]{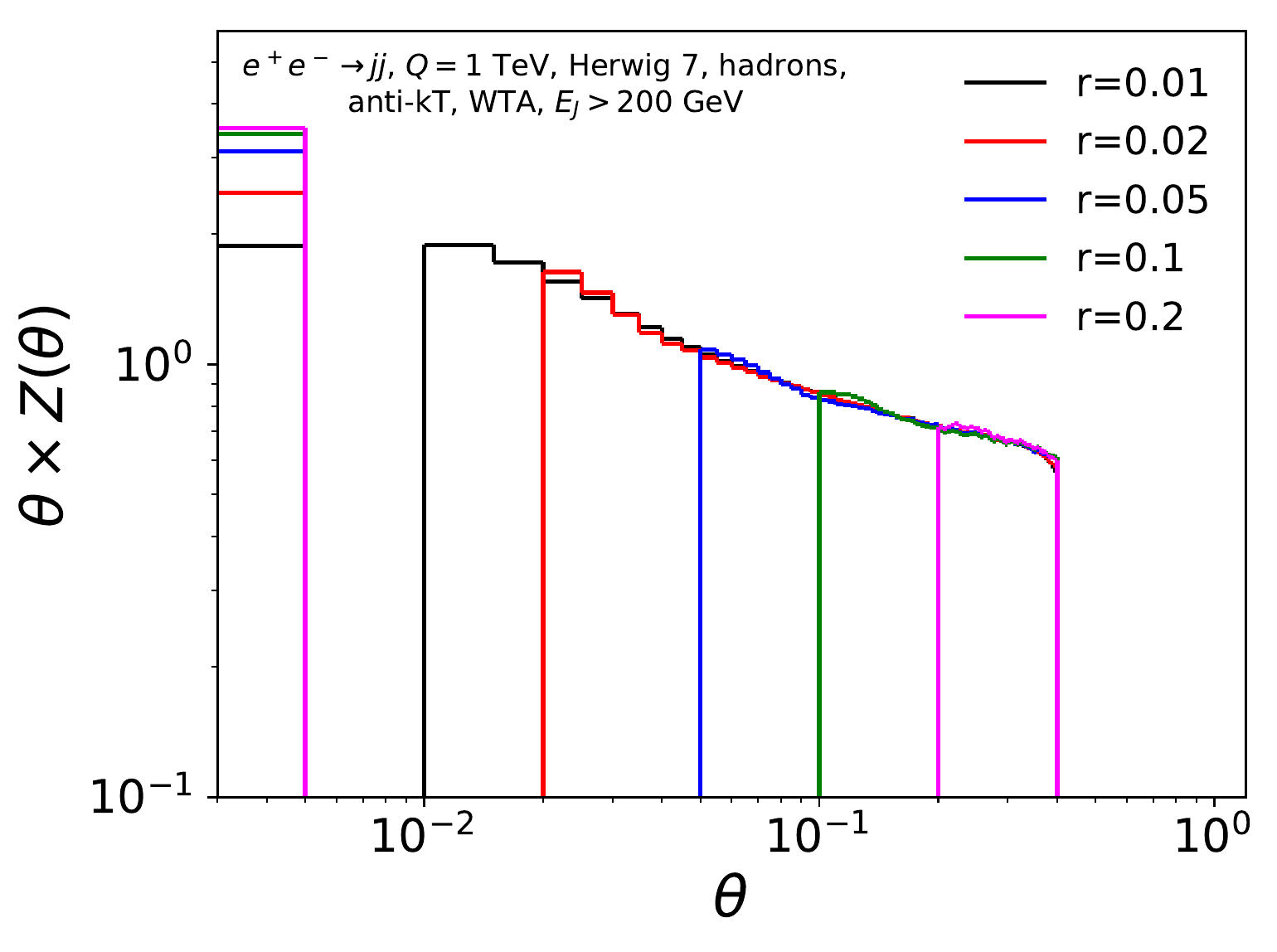}
\includegraphics[width=0.328\textwidth]{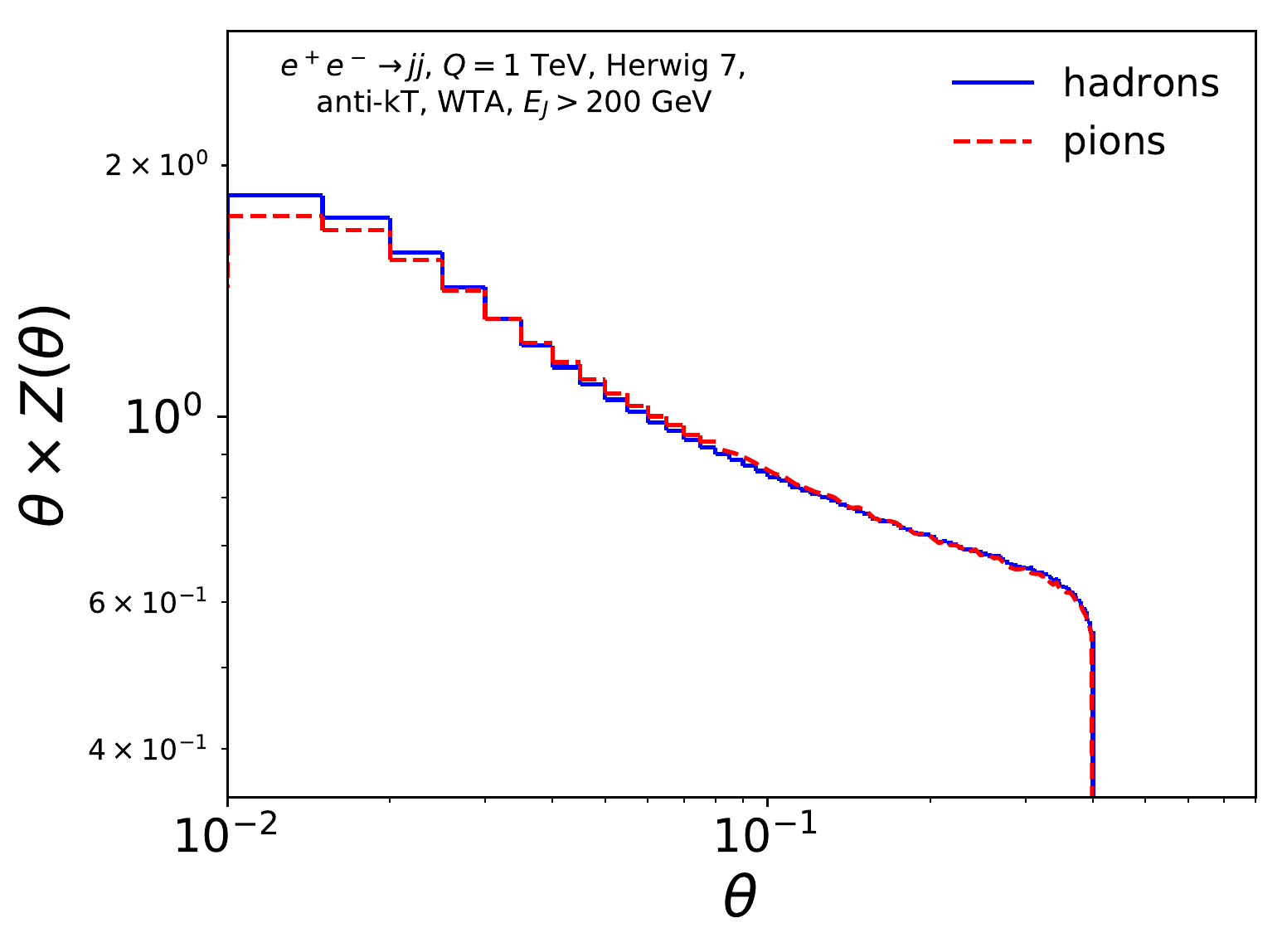}%
\includegraphics[width=0.328\textwidth]{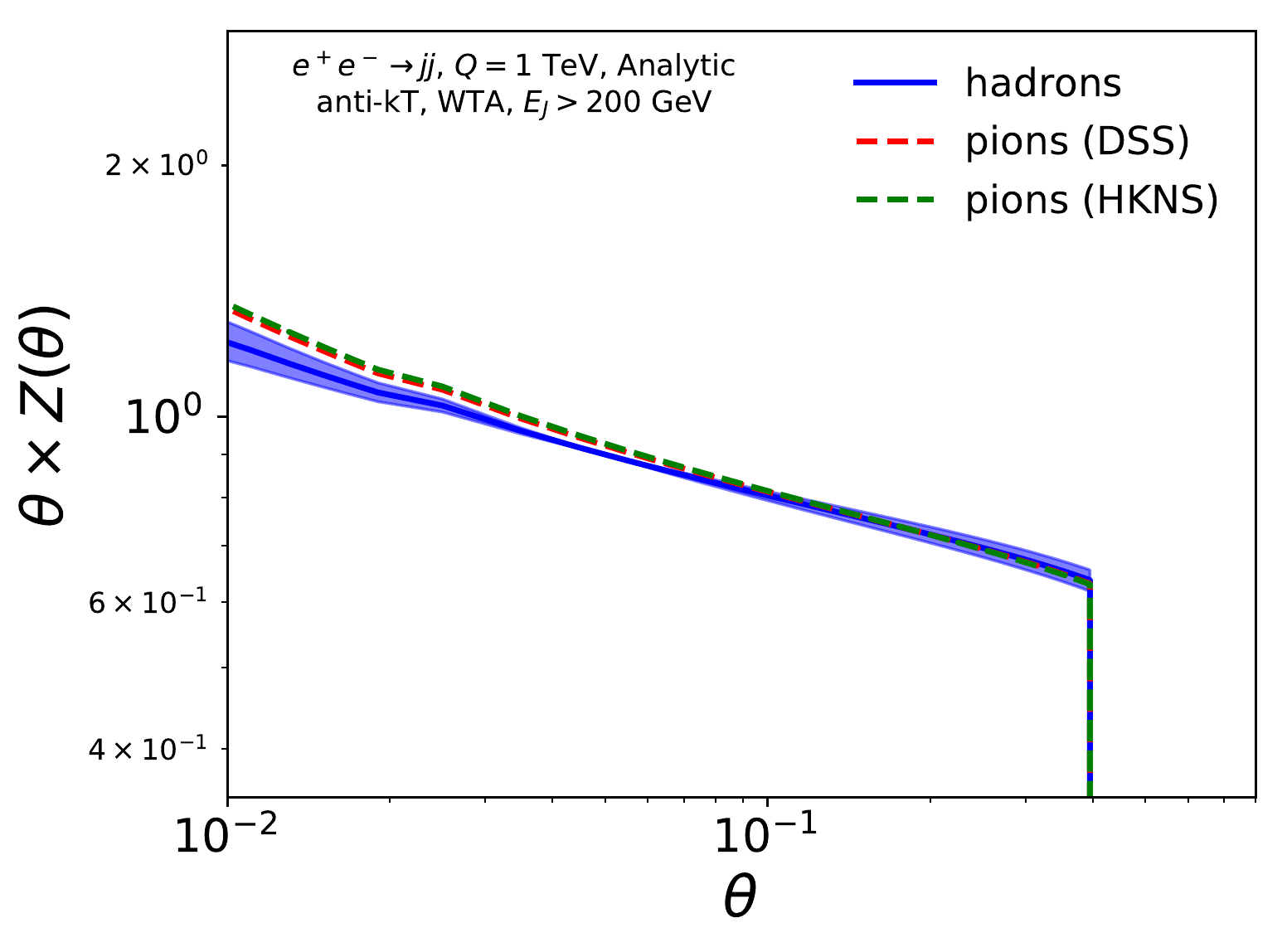}%
\caption{Left panel: The jet shape using subjets (instead of hadrons)
  as input. The lowest bin gets its contribution from $\theta=0$. All the jet shapes were normalized by the same factor, taken to be the area of the $r=0.01$ curve in $\theta = [0.1, 0.4]$. 
   Middle panel: The jet shape using all particles (blue) and
  only charged pions (red dashed) as input. The red-dashed
  pion distribution has been normalized to the blue curve for all particles
  to emphasize the similarity in the shape. Right panel: The analytic
  calculation of the jet shape for pions using two set of
  fragmentation functions: DSS or HKNS (red and green dashed
  respectively), compared to the full result (blue).}
\label{fig:zint_subjets}
\end{figure}

The attentive reader will have noticed that our jet shape distributions go down to angles smaller than the size of a calorimeter cell at the LHC ($\theta \approx 0.1$). In the left panel of \fig{zint_subjets} we demonstrate that limited angular resolution does not change the jet shape. Specifically, we recluster the jet into subjets of radius $r < R$, and then calculate the jet shape using these subjets (instead of hadrons) as input. The distributions overlap as long as the angle is above the subjet radius scale, which is expected from our calculations in \sec{framework}. Below the subjet radius the distribution drops off, except for the contribution from $\theta=0$. Alternatively, a more granular angular resolution can be achieved using tracking, so we show in the right panel of \fig{zint_subjets} that the jet shape defined on all particles or only on charged pions has the same shape. Of course this distribution is a bit more sensitive to hadronization effects, and requires few nonperturbative parameters to implement in our analytic calculation, as discussed in \sec{jetshape}.

\begin{figure}[t]
\centering
\includegraphics[width=0.4\textwidth]{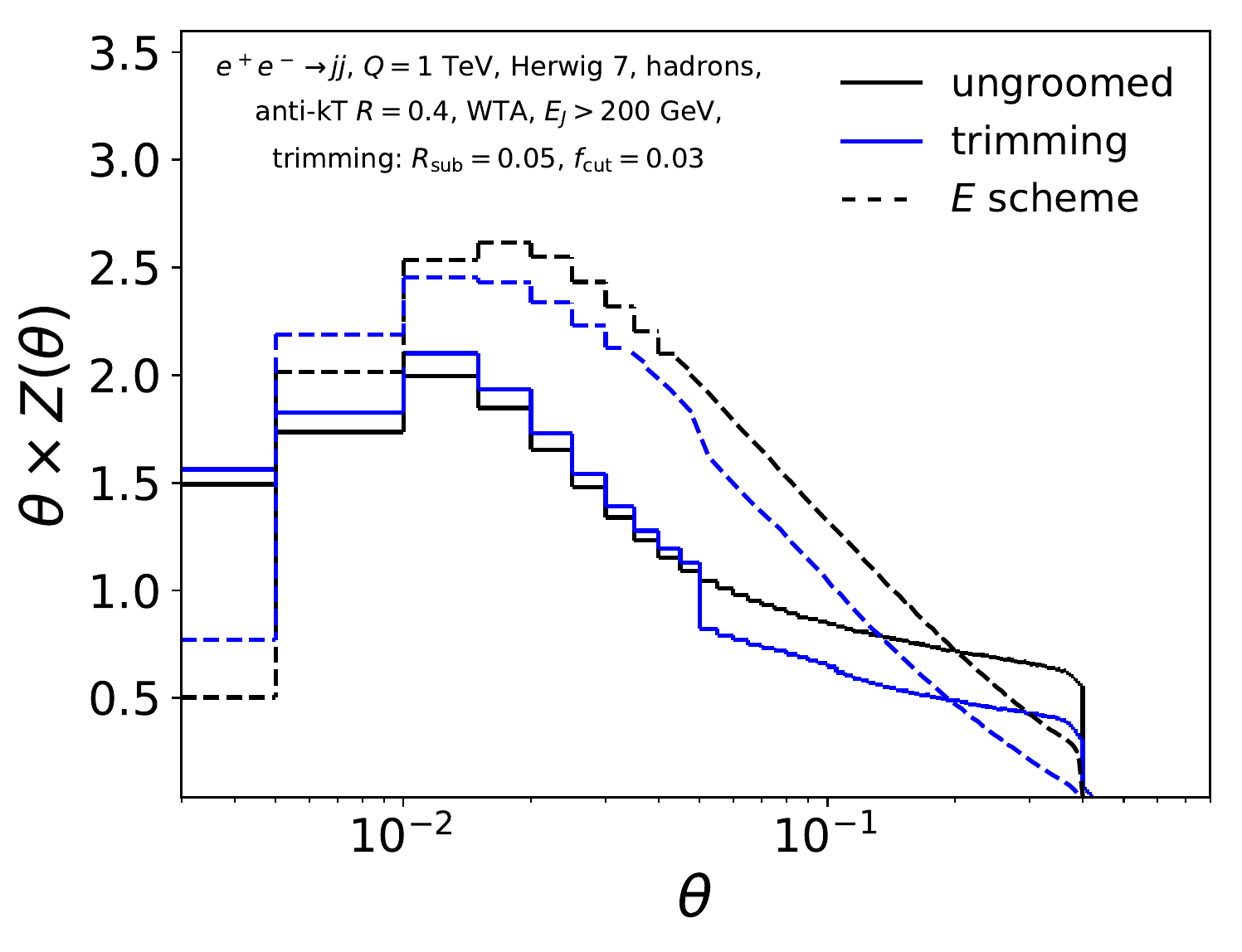}
\includegraphics[width=0.4\textwidth]{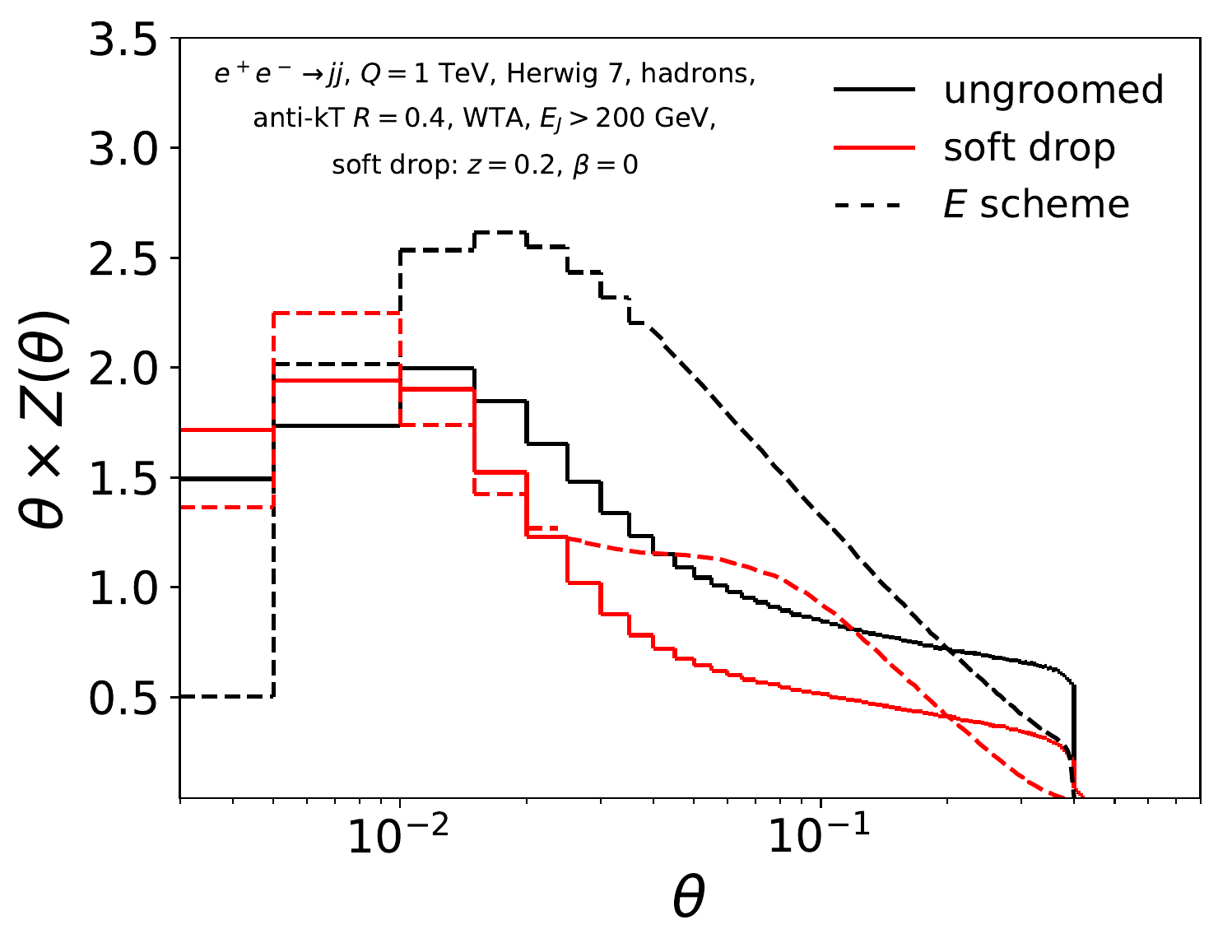}
\caption{The jet shape with respect to the WTA axis (solid) and
  standard jet axis (dashed) at hadron level.
  Left panel:
  after applying trimming (blue). Right panel: after soft drop (red).}
\label{fig:zint_groom}
\end{figure}

To investigate how sensitive the jet shape is to soft radiation, we aggressively\footnote{This is necessary to see any effect, due to the limited amount of radiation in the $e^+e^-$ environment.} remove soft radiation using a grooming procedure. Specifically, we consider trimming~\cite{Krohn:2009th} with $R_{\rm sub} = 0.05$ and $f_{\rm cut} = 0.03$, and soft drop~\cite{Larkoski:2014wba} with $z_\mathrm{cut}=0.2$ and $\beta = 0$. For comparison we show results both for the jet shape with the WTA axis and the standard ($E$ scheme) jet axis, see \fig{zint_groom}. We take the momentum fraction $z$ to be the hadron energy divided by the groomed jet energy, but have normalized the distributions to those of ungroomed jet, so one can clearly see how much radiation is removed by the grooming. For trimming there is little change for $r < 0.05$, since the subjet containing the WTA axis is never trimmed away. For $r>0.05$ an almost constant amount is removed by trimming. By contrast, the jet shape for the standard jet axis affects all angles, due to the response of the axis to the trimming. Similarly, soft drop removes a constant amount, except close to het jet axis. 

\begin{figure}[t]
\centering
\includegraphics[width=0.4\textwidth]{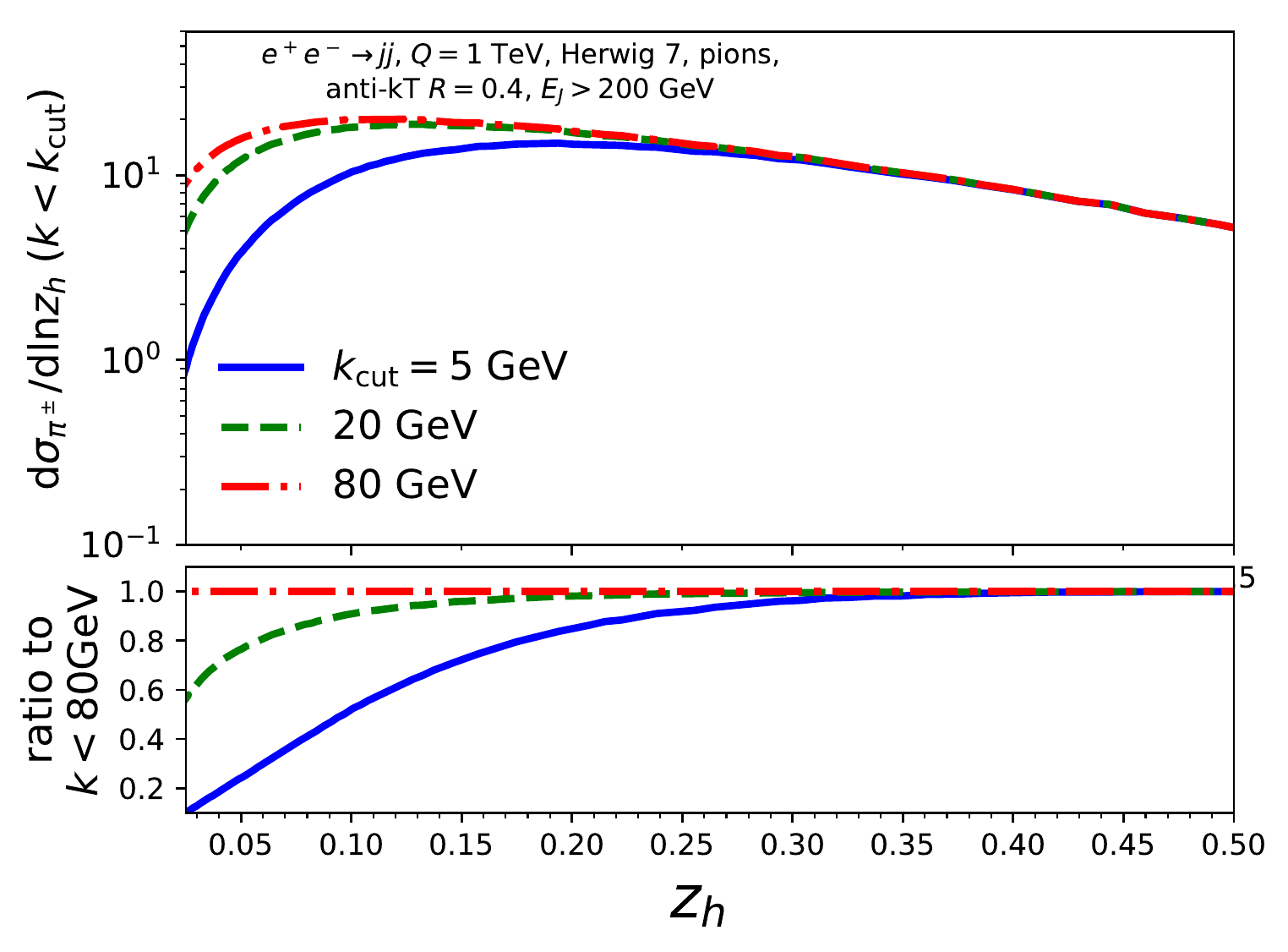}
\includegraphics[width=0.4\textwidth]{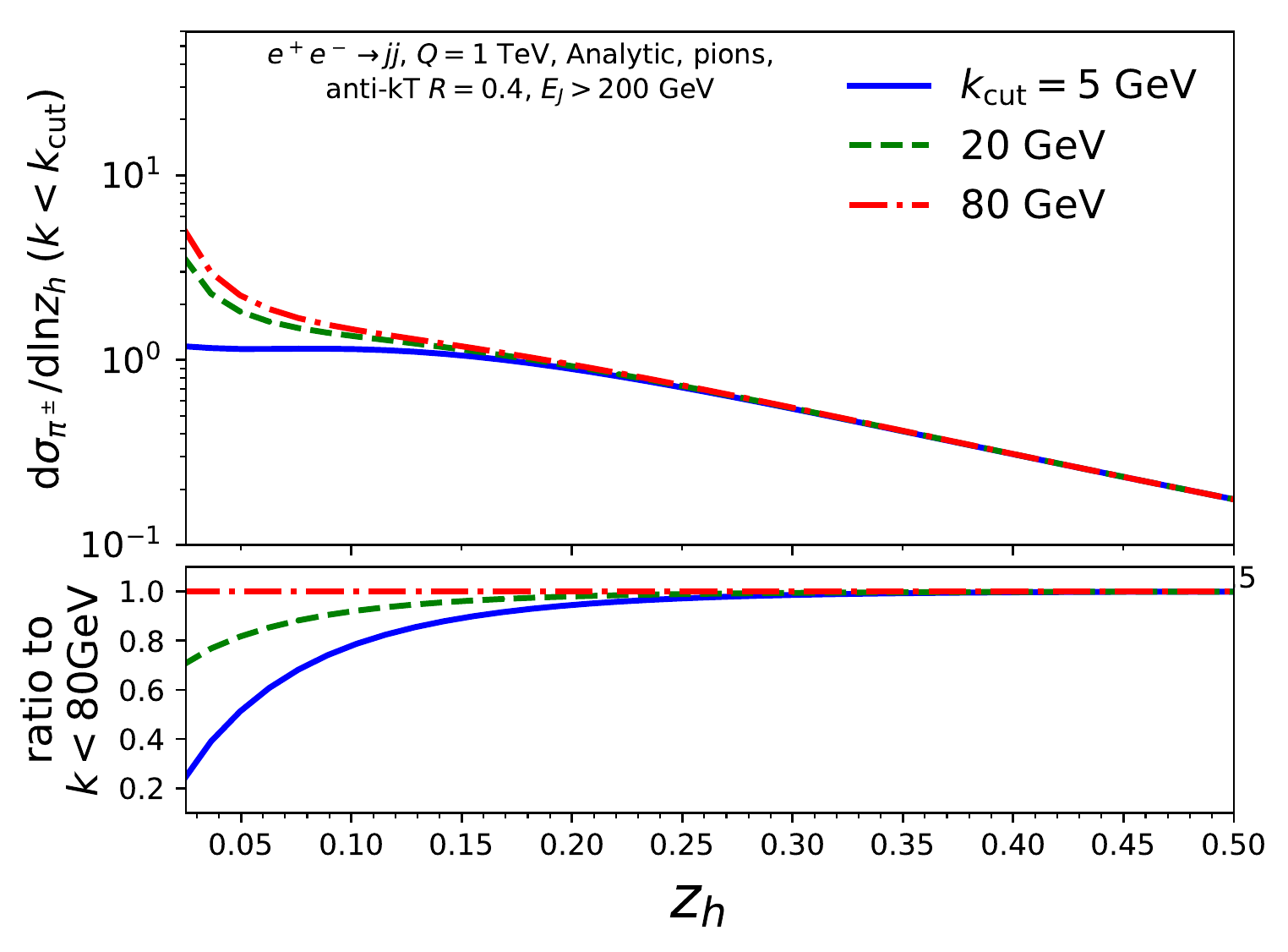} 
\includegraphics[width=0.4\textwidth]{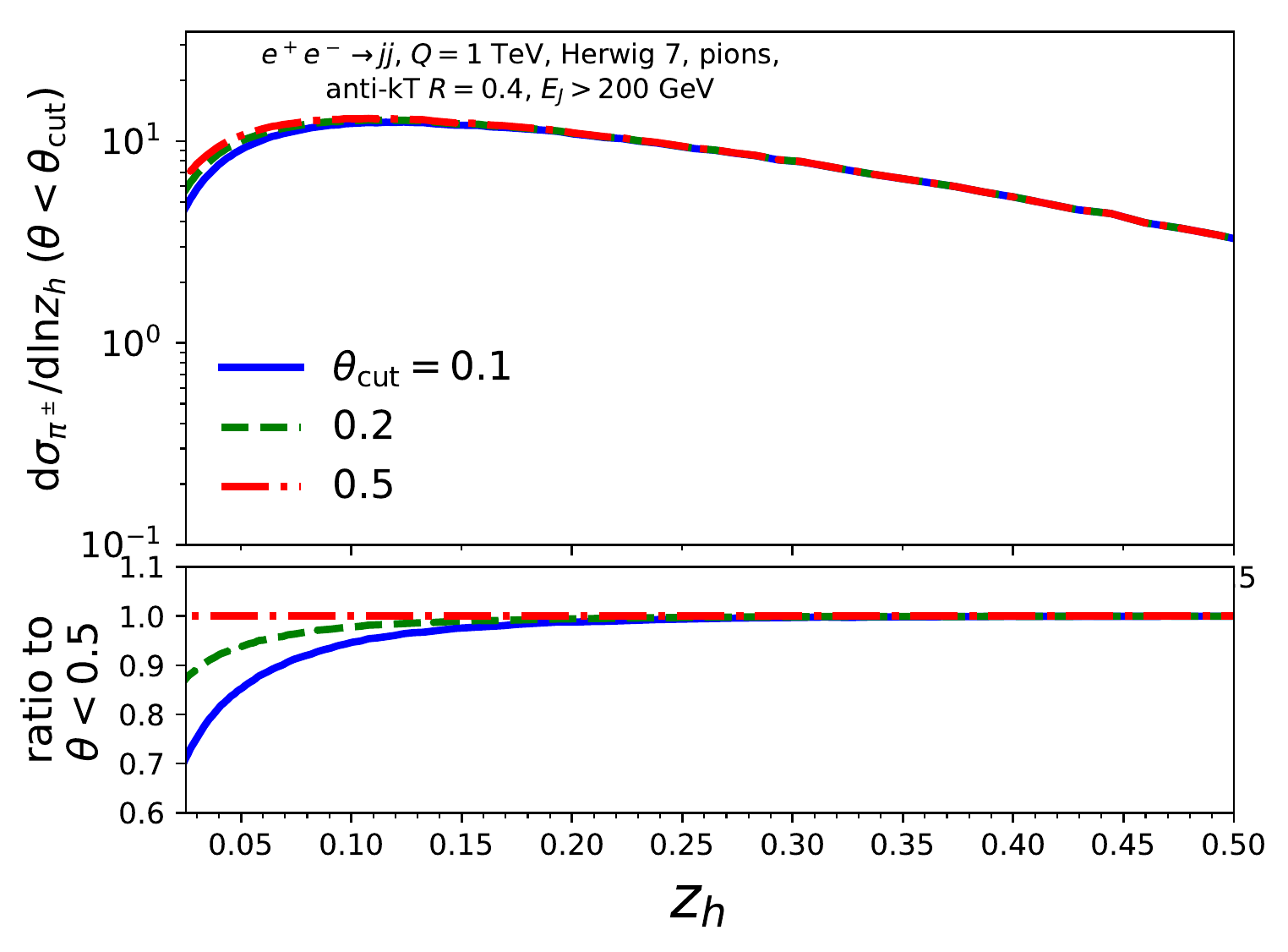}%
\includegraphics[width=0.4\textwidth]{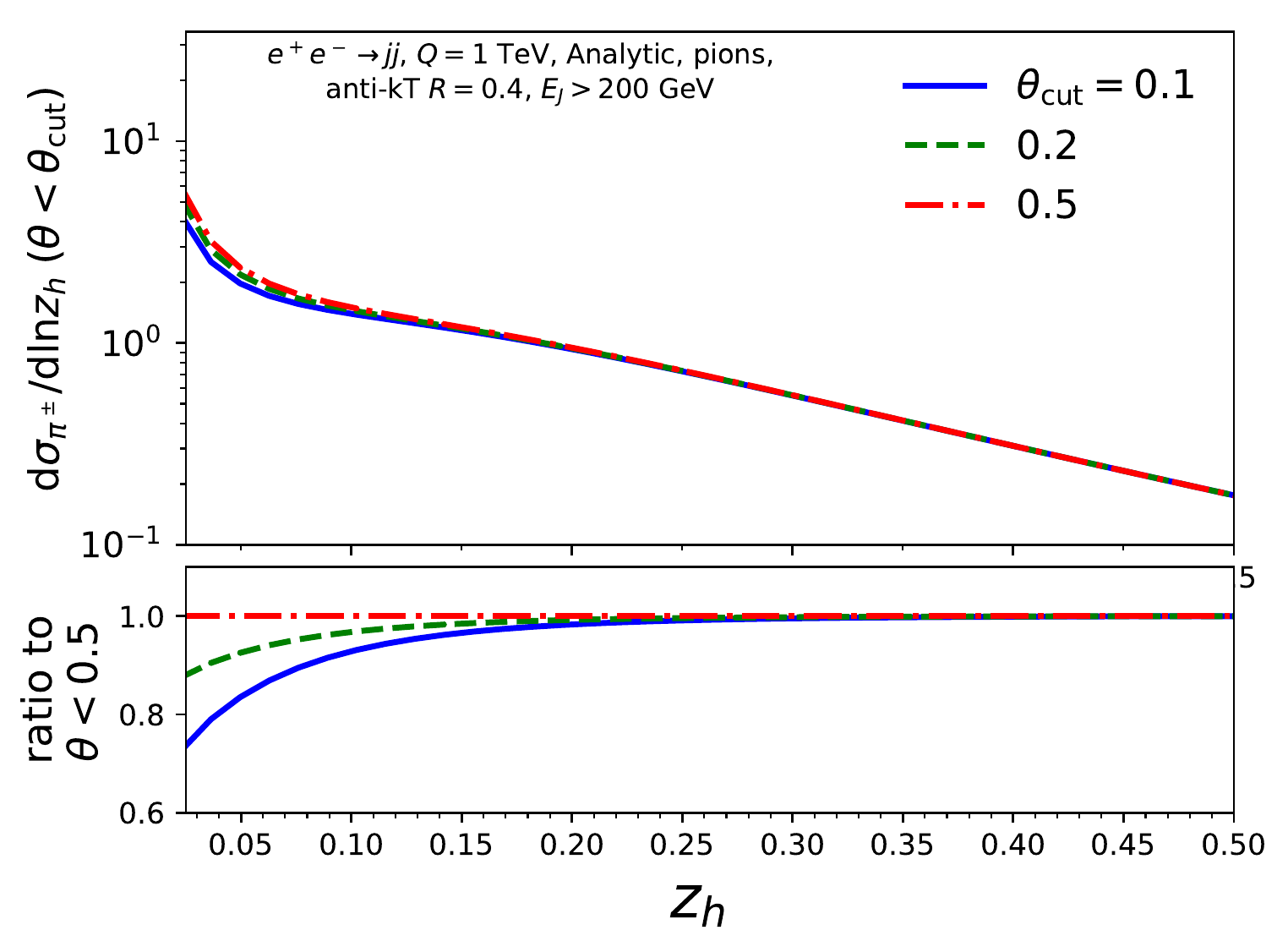}%
\caption{The fragmentation spectrum of charged pions with a cut on their transverse momentum (top) or angle (bottom), from Monte Carlo (left) and through the analytic calculation (right). The ratio to the ``most-inclusive'' cuts, $\theta_{\rm cut} = 0.5$ or $k_{\rm cut} = 80$~GeV, are shown for each case.}
\label{fig:zplot_pions}
\end{figure}

We next consider the effect of a cut on transverse momentum or angle on the fragmentation spectrum of charged pions, see \fig{zplot_pions}. The effect of these cuts is clearly visible at small $z_h$ but does not affect the distribution at large $z_h$, since such hadrons are kinematically forced to be close to the jet axis. In particular, for $z_h>0.5$ the winner-take-all axis is along the hadron and the distribution is insensitive to the cut, which is why we did not show this region.
The analytic curves exhibit the same qualitative behavior as those obtained using \Herwig, as is particularly clear in the subpanels which show the ratio to the ``most-inclusive'' cuts on either $\theta$ or $k$. The absolute distributions differ, but this is simply indicative of the different intrisically-nonperturbative fragmentation spectrum used by \Herwig and DSS. 

\begin{figure}[t]
\centering
\includegraphics[width=0.45\textwidth]{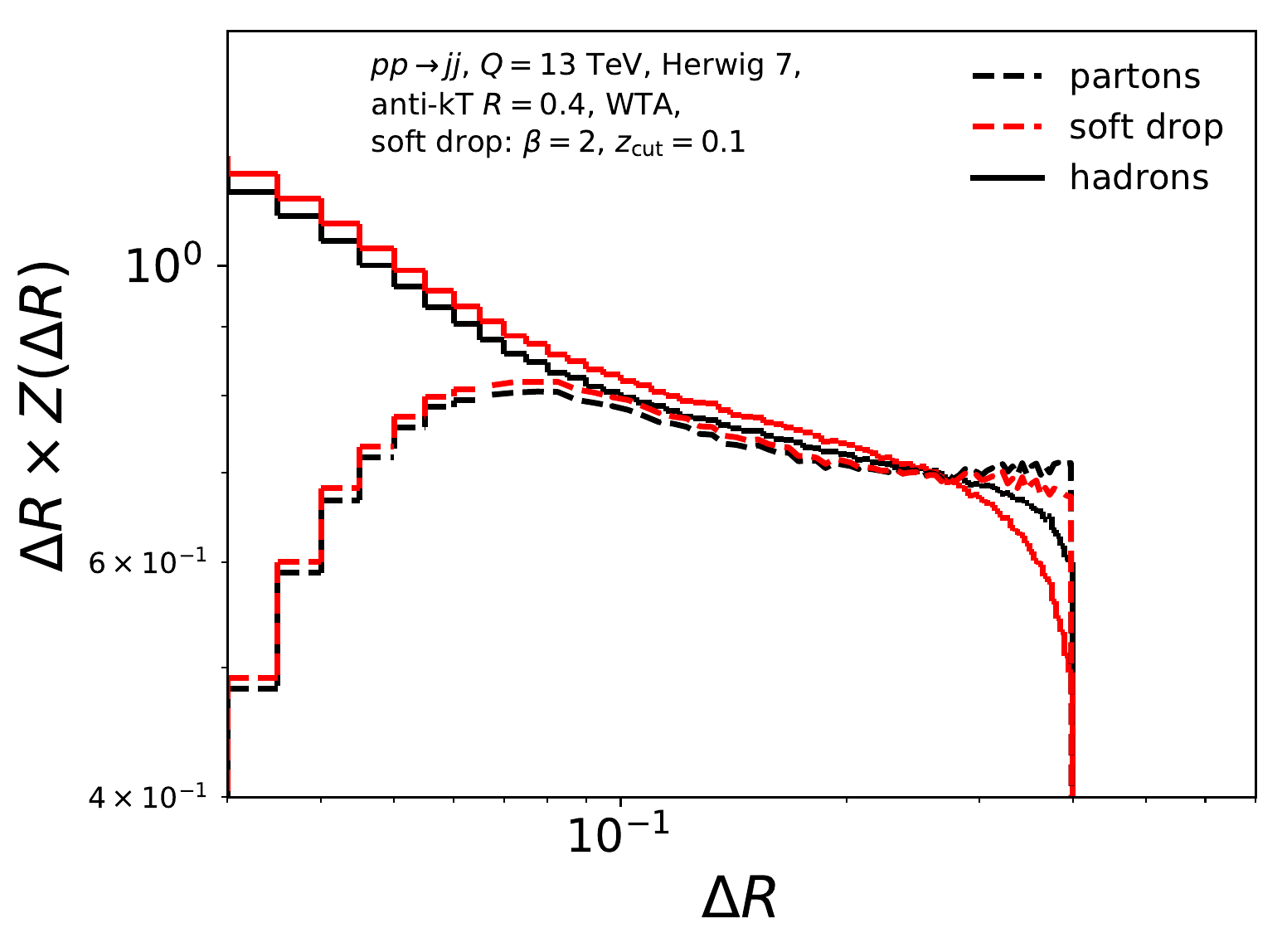}
\caption{Jet shape at the LHC for jets with $p_{T,J}> 200$ GeV at hadron (solid) and parton level (dashed), using all particles (black) or after applying soft drop (red).}
\label{fig:zint_lhc}
\end{figure}

We end this section by presenting results for the jet shape at the 13~TeV LHC. The simulated events are clustered using the standard ($pp$-version) of anti-$k_T$, and the jets are required to have transverse momentum $p_{T,J} > 200$ GeV with respect to the beam axis. We now take $z_h = p_{T,h} / p_{T,J}$, i.e.~the  transverse momentum fraction, and use $\Delta R = \sqrt{(\phi_h \!-\! \phi_J)^2 + (y_h \!-\! y_J)^2}$ to quantify the distance to the jet axis, where $\phi$ and $y$ are the azimuthal angle and rapidity. Specifically, we show
\begin{align}
Z(\Delta R) = 
\Big(\frac{\df \langle z \rangle}{\df \Delta R}\Big)\Big/\Big(\int_{\Delta R_{\rm min}}^R\! \df \Delta R\,\frac{\df \langle z \rangle}{\df \Delta R}\Big)
\,,\end{align}
in \fig{zint_lhc}, where $\Delta R_{\rm min} = 0.1$. 
It's clear that the jet shape exhibits the same (approximate) power-law behavior observed for the $e^+e^-$ case. Also shown is the jet shape after soft drop with $z_\mathrm{cut}=0.1$ and $\beta = 2$. This grooming is less aggressive than in \fig{zint_groom} and only affects the region close to the jet boundary.

\begin{figure}[t]
\centering
\includegraphics[width=0.45\textwidth]{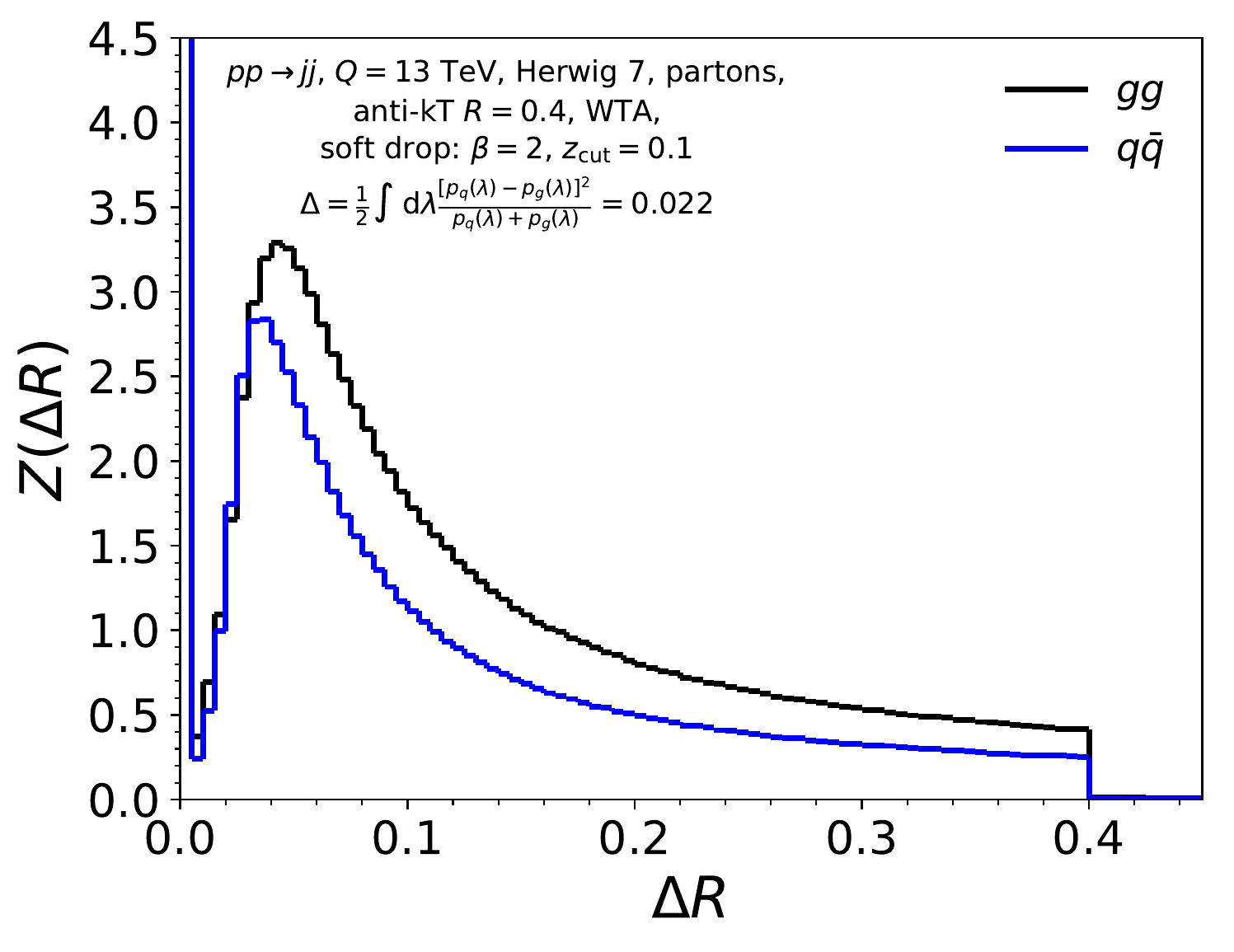}
\caption{Jet shape at the LHC for jets with $p_{T,J}> 200$ GeV at parton level for $pp \to q\bar{q}$ (blue) and $pp \to g g$ (black). Note that the curves have been normalized in the whole range, i.e. with $\Delta R_{\rm min} = 0.0$.}
\label{fig:zint_qg}
\end{figure}

Last of all we show in \fig{zint_qg} the jet shape in $pp$ collisions for quark vs.~gluon jets which we define as being produced by the tree-level hard-scattering process $pp \to q \bar q$ vs.~$pp \to gg$. The quark distribution lies below the gluon distribution, except for the first bin where it is much higher, since both curves have the same normalization (this time with $\Delta R_{\rm min} = 0.0$). As expected, the gluon distribution is broader, since gluons radiate more than quarks. We also calculate the classifier separation, 
\begin{align} \label{eq:classifier}
 \Delta = \frac12 \int\! \df \lambda\, \frac{\big[p_q(\la) - p_g(\la)\big]^2}{p_q(\la)+p_g(\la)}
\,,\end{align}
and find that for the WTA jet shape $\Delta = 0.022$. Compared to the observables studied in refs.~\cite{Badger:2016bpw, Gras:2017jty}, this is not a particularly powerful quark-gluon discriminant.

\section{Conclusions}
\label{sec:conc}

In this paper we have studied the distribution of hadrons inside a jet in terms of their energy and angle with respect to a recoil-free axis. Instead of the usual double-logarithmic dependence of the cross section on the angle (or transverse momentum), we find a power law, because these observables are insensitive to soft radiation. Since the position of the axis is not smeared by soft radiation, these observables are particularly interesting to study perturbative \emph{collinear} physics, which feature prominently in our distributions when approaching the axis. 

In addition to the intrinsic interest in TMD fragmentation of the nuclear physics community, we believe these observables are promising for studying the quark-gluon plasma, since the medium produces so much low-energy radiation that it is essential to use an axis that is insensitive to that. Other potential applications include the extraction of the strong coupling $\al_s$ or the discrimination of quark and gluon jets at the LHC. 
What makes these observables interesting from a theoretical point of view is that they are purely collinear and so they can be calculated to higher orders from the collinear splitting functions. Another direction we intend to explore is to consider recoil-free axes for more complicated observables, e.g.~those used to tag boosted heavy objects.

\begin{acknowledgments}
This work is supported by the US Department of Energy contract DE-AC52-06NA25396, the LANL/LDRD Program, the ERC grant ERC-STG-2015-677323, and the D-ITP consortium, a program of the Netherlands Organization for Scientific Research (NWO) that is funded by the Dutch Ministry of Education, Culture and Science (OCW). 
\end{acknowledgments}

\phantomsection
\addcontentsline{toc}{section}{References}
\bibliographystyle{jhep}
\bibliography{wta_phen}

\end{document}